\documentclass[twocolumn,showpacs,aps,superscriptaddress]{revtex4-1}
\usepackage{amsmath}
\usepackage[switch, modulo]{lineno}
\usepackage{graphicx}
\usepackage{longtable}
\usepackage{ltablex}
\include{def}
\def\Journal#1#2#3#4{{#1} {#2} (#3) #4}

\def\JPG{J.Phys.G}
\def\JHEP{JHEP}
\def\EPJ{Eur.Phys.J.}
\def\ARNPS{Ann.Rev.Nucl.Part.Sci.}
\def\NPA{Nucl.Phys.A}
\def\NPB{Nucl.Phys.B}

\def\PLB{Phys.Lett.B}

\def\PRL{Phys.Rev.Lett.}
\def\PRC{Phys.Rev.C}
\def\PRD{Phys.Rev.D}
\def\PR{Phys.Rev.}

\def\EPJ{Eur.Phys.J.}

\def\APP{Acta Phys. Pol.}
\def\be{\begin{equation}}
\def\ee{\end{equation}}

\begin{document}

\title{
Geometrical scaling for energies available at the BNL Relativistic
    Heavy Ion Collider to those at the CERN Large Hadron Collider\\
}

\author{M.Petrovici}
\affiliation{National Institute for Physics and Nuclear Engineering - IFIN-HH\\
Hadron Physics Department\\
Bucharest - Romania}
\affiliation{Faculty of Physics, University of Bucharest}
\author{A.Lindner}
\affiliation{National Institute for Physics and Nuclear Engineering - IFIN-HH\\
Hadron Physics Department\\
Bucharest - Romania}
\affiliation{Faculty of Physics, University of Bucharest} 
\author{A.Pop} 
\affiliation{National Institute for Physics and Nuclear Engineering - IFIN-HH\\
Hadron Physics Department\\
Bucharest - Romania}
\author{M.T{\^ a}rzil{\u a}}
\affiliation{National Institute for Physics and Nuclear Engineering - IFIN-HH\\
Hadron Physics Department\\
Bucharest - Romania}
\affiliation{Faculty of Physics, University of Bucharest} 
\author{I.Berceanu}
\affiliation{National Institute for Physics and Nuclear Engineering - IFIN-HH\\
Hadron Physics Department\\
Bucharest - Romania}

\date{\today}

\begin{abstract}

   Based on the recent RHIC and LHC experimental results, the $\langle p_T\rangle$
dependence of identified light flavour charged hadrons on $\sqrt{(\frac{dN}{dy})/S_{\perp}}$, relevant scale in gluon saturation 
picture, is studied from $\sqrt{s_{NN}}$=7.7 GeV up to 5.02 TeV. This study is extended to the slopes of 
the $\langle p_T\rangle$ dependence on the particle mass
and the $\langle\beta_T\rangle$ parameter from Boltzmann-Gibbs Blast Wave (BGBW) fits 
of the $p_T$ spectra. 
A systematic decrease of the slope of the $\langle p_T\rangle$ dependence on $\sqrt{(\frac{dN}{dy})/S_{\perp}}$
from BES to the LHC energies is evidenced. While for the RHIC energies, within the experimental
errors, the 
$\langle p_T\rangle$/$\sqrt{(\frac{dN}{dy})/S_{\perp}}$ does not depend on centrality, at the LHC energies a deviation from a linear behaviour is observed towards the most central collisions.
The influence of the corona contribution to the observed trends is discussed. 
The slopes of the $\langle p_T\rangle$ particle mass dependence and the
$\langle\beta_T\rangle$ parameter from BGBW fits scale well with $\sqrt{(\frac{dN}{dy})/S_{\perp}}$.
Similar systematic trends for pp at $\sqrt{s}$=7 TeV are in a good agreement with the ones corresponding
to Pb-Pb collisions at $\sqrt{s_{NN}}$=2.76~TeV and 5.02 TeV pointing to a system size independent behaviour.

\end{abstract}


\maketitle

\section{Introduction}

  Parton density evolution as a function of x and $Q^2$, addressed more than 35 years
ago \cite{GRIB1} and its experimental confirmation at HERA 
\cite{HERA1} have triggered a real interest in the community studying 
ultra-relativistic heavy ion collisions. The rise of the structure function at 
low x, still visible at small values of $Q^2$ \cite{HERA2, HERA3} where the
perturbative QCD does not work anymore, requires new approaches for a complete
understanding 
of the \mbox{log$\frac{1}{x}$-log$Q^2$} QCD landscape. Low x values and moderate $Q^2$
are characteristic features for the early stage of hadron collisions starting from 
RHIC up to LHC energies. For average transverse momentum ($\langle p_T\rangle$) values of the order of 1-2 GeV/c, specific 
for this range of energies, the x values at mid rapidity are of the order of 
\mbox{$\sim 10^{-2}$} and \mbox{$\sim 10^{-4}$} respectively. Such initial conditions are used by 
different theoretical approaches for describing especially the most recent results 
from LHC energies. Colour Glass Condensate (CGC) is one of such approaches based 
on strong classical colour fields description of the small x degrees of freedom
\cite{McL1, Ianc1, KLN}. Local parton-hadron duality picture (LPHD) \cite{LPHD} and 
dimensionality argument \cite{Lev1, Lap1} predict a decrease of 
the ratio between the average transverse momentum  and the square root of the hadron 
multiplicity per unit of rapidity and unit of the colliding hadrons transverse overlapping area 
($\langle p_T\rangle$/$\sqrt{(dN/dy)/S_{\perp}}$) towards central collisions and 
higher energies. With the latest results from the Beam Energy Scan (BES) at 
RHIC and the highest energies at LHC, it is worth revisiting such a dependence. Recently
evidenced similarities between pp, p-Pb and Pb-Pb collisions at LHC energies 
in terms of long range near side two particle correlations, transverse flow and
strangeness enhancement as a function of charged particle multiplicity 
\cite{CMS1, CMS2, ATLAS1, ALICE9, Cristi1, LHCb1, ALICE10} support the idea that even in small 
colliding systems, due to increased parton density at such energies, the probability
of multiple parton interaction increases, the rescattering processes become important 
and a thermalised stage could be reached although the interaction time is extremely short.
Such a high density deconfined small system could follow a hydrodynamic type expansion. 
To what extent the hydrodynamics is applicable in small systems is still under debate \cite{Shu}. The most successful phenomenological models,
UrQMD, HIJING, NeXSPheRIO, AMPT, PHSD, EPOS, 
describing the latest results 
obtained at LHC in pp, p-Pb and Pb-Pb collisions are based on combinations of different 
approaches for different stages of the collision \cite{UrQMD, HIJING, NeXSPheRIO, HIJING, AMPT, PHSD, EPOS} 
while
the classical phenomenological models used in particle physics like PYTHIA 
\cite{PYTHIA}, HERWIG \cite{HERWIG} and PHOJET \cite{PHOJET} 
had to implement processes like multiparton interaction, rescattering, colour 
reconnection \cite{CR} or shoving mechanism \cite{Shov} in order to improve the 
agreement with the LHC results, especially in the soft sector in pp collisions.          
In this paper we also present a comparison between pp and Pb-Pb at LHC energies in
terms of the dependence of different observables on the
$\sqrt{(dN/dy)/S_{\perp}}$ variable.
 In the second chapter of the paper the estimates of the overlapping
area of the colliding hadrons are presented. Details on the hadron density per unit of rapidity  are given in the third chapter. The $\langle p_T\rangle$ dependence on 
$\sqrt{(dN/dy)/S_{\perp}}$ is presented in Chapter IV for BES and $\sqrt{s_{NN}}$=62.4,
130, 200 GeV Au-Au collisions measured by the STAR Collaboration at RHIC and for Pb-Pb
collisions at $\sqrt{s_{NN}}$=2.76 and 5.02 TeV measured by
\linebreak
the ALICE Collaboration at LHC. Chapter V is dedicated to the
$\sqrt{(dN/dy)/S_{\perp}}$ dependence of the slope of the linear $\langle p_T\rangle$ 
versus 
particle mass behaviour for identified light flavour charged hadrons. The BGBW fit parameters of 
$p_T$ spectra are presented versus the same geometrical variable of the colliding systems
in Chapter VI. Similarities, in terms of $\sqrt{(dN/dy)/S_{\perp}}$ dependence of
different observables, in pp at $\sqrt{s}$=7 TeV and Pb-Pb at $\sqrt{s_{NN}}$=2.76 and 5.02 TeV are discussed in Chapter VII.
Chapter VIII is dedicated to conclusions. 
\vspace{-1.00cm}
\section{Overlapping area $S_{\perp}$ estimates}
   The overlapping area of the two colliding nuclei for a given incident energy and centrality was estimated
based on the Glauber Monte Carlo (GMC) approach \cite{Gla1, Gla2, Gla3, Glis}. For the nuclear density profile of the colliding
nuclei, a Wood-Saxon distribution was considered: 
\begin{equation}
\rho(r)=\frac{1}{1+exp(\frac{r-r_0}{a})}
\end{equation}
with a=0.535 fm, $r_0$=6.5 fm for the Au nucleus \cite{STAR1} and a=0.546 fm, $r_0$=6.62 fm for the Pb nucleus \cite{ALICE1}.
Within the black disc approach, the nucleons are considered to collide if the relative transverse distance 
\mbox{d $\leq$ $\sqrt{\frac{\sigma_{pp}}{\pi}}$}, where $\sigma_{pp}$ is the nucleon-nucleon interaction cross section. 
The $\sigma_{pp}$ values for the corresponding $\sqrt{s_{NN}}$ energies were taken from \cite{AMAL1, STAR1, ALICE1,ALICE21}.
The main characteristics of the collision at different centralities
for Au-Au at 
$\sqrt{s_{NN}}$=7.7, 11.5, 19.6, 27 and 39 GeV obtained in the Beam Energy Scan (BES) at RHIC \cite{STAR2}, Au-Au at $\sqrt{s_{NN}}$=62.4, 130 and 200 GeV \cite{STAR1} and Pb-Pb at 
$\sqrt{s_{NN}}$=2.76 and 5.02~TeV \cite{ALICE1, ALICE21} are presented in Table I
(see caption for notations).
The geometrical overlapping areas ($S_{\perp}^{geom}$) have been estimated by averaging the maximum values of the x and y coordinates determined per event, over many events. $S_{\perp}^{var}$ has been estimated as being proportional 
to the quantity \mbox{S=$\sqrt{<\sigma_x^2><\sigma_y^2>-<\sigma_{xy}>^2}$},
$\sigma_x^2$, $\sigma_y^2$ are the variances and $\sigma_{xy}$ is the co-variance
of the participant distributions in the transverse plane, per event
\cite{Alv}. They were
averaged ($<...>$) over many events.
\begin{longtable*}{|c|c|c|c|c|c|c|c|c|c|}
\hline
\bf System & $\bf \sqrt{s_{NN}}$ & \bf Cen.&  $\bf <N_{part}>$ &  
$\bf S_{\perp}^{geom} $ & $\bf S_{\perp}^{var}$ & $ \bf f_{core}$ & $\bf (S_{\perp}^{geom})^{core} $& $\bf (S_{\perp}^{var})^{core} $ & \bf dN/dy \\ 
 &  (GeV) & ($\%$) &  & $(fm^2)$ & $(fm^2)$ & & $(fm^2)$ & $(fm^2)$ &\\ 
\hline
       &   &  0-5  & 337$\pm$2 & 146.1$\pm$0.7 & 147.1$\pm$0.7& 0.88$\pm$0.00 & 126.5$\pm$0.6&  124.6$\pm$0.6 & 476.7$\pm$22.5\\
       &  &  5-10 & 290$\pm$6 & 126.9$\pm$0.7 & 129.7$\pm$0.6& 0.84$\pm$0.00 & 107.6$\pm$0.7& 105.0$\pm$0.5 & 392.5$\pm$18.5\\
	   &  & 10-20 & 226$\pm$8 & 103.6$\pm$0.7 & 108.9$\pm$0.5& 0.80$\pm$0.00 & 85.5$\pm$0.7& 84.3$\pm$0.4 & 295.4$\pm$14.0\\
  &  & 20-30 & 160$\pm$10& 79.7$\pm$0.8 & 87.1$\pm$0.4& 0.75$\pm$0.00 & 63.3$\pm$0.7& 63.8$\pm$0.3  & 203.6$\pm$9.8\\
Au-Au	   & 7.7 & 30-40 & 110$\pm$11& 61.2$\pm$0.8 & 70.4$\pm$0.3& 0.70$\pm$0.00 & 46.6$\pm$0.7&  48.9$\pm$0.2 & 135.1$\pm$6.4\\
	   &  & 40-50 & 72$\pm$10 & 45.9$\pm$0.8 & 56.9$\pm$0.3& 0.63$\pm$0.00 & 33.1$\pm$0.8&  37.5$\pm$0.2 & 84.8$\pm$4.1\\
	   &  & 50-60 & 45$\pm$9  & 32.8$\pm$0.8 & 45.7$\pm$0.2& 0.56$\pm$0.00 & 21.9$\pm$0.8&  28.9$\pm$0.1 & 51.5$\pm$2.5\\
	   &  & 60-70 & 26$\pm$6  & 21.8$\pm$0.8 & 36.3$\pm$0.2& 0.47$\pm$0.00 & 12.8$\pm$0.7&  22.6$\pm$0.1 & 27.6$\pm$1.4\\
	   &  & 70-80 & 14$\pm$4  & 12.1$\pm$0.7 & 26.8$\pm$0.2& 0.37$\pm$0.00 & 5.4$\pm$0.6&  15.6$\pm$0.1 & 13.8$\pm$0.8\\
	   \hline 
      &    &  0-5  & 338$\pm$2  & 146.1$\pm$0.7 & 147.1$\pm$0.7 & 0.88$\pm$0.00 & 126.5$\pm$0.6 & 124.5$\pm$0.6 & 565.1$\pm$29.5\\
       &  &  5-10 & 291$\pm$6  & 126.6$\pm$0.7 & 129.5$\pm$0.6 & 0.84$\pm$0.00 & 107.2$\pm$0.7 & 104.6$\pm$0.5 & 449.0$\pm$23.3\\
	   &  & 10-20 & 226$\pm$8  & 103.5$\pm$0.7 & 108.9$\pm$0.5 & 0.80$\pm$0.00 & 85.3$\pm$0.7 & 84.1$\pm$0.4 & 335.5$\pm$17.5\\
  &  & 20-30 & 160$\pm$9  & 79.9$\pm$0.8 & 87.3$\pm$0.4 & 0.75$\pm$0.00 &  63.4$\pm$0.7 & 63.9$\pm$0.3 & 225.5$\pm$11.8\\
Au-Au	   & 11.5 & 30-40 & 110$\pm$10 & 61.3$\pm$0.8 & 70.5$\pm$0.3 & 0.70$\pm$0.00 & 46.6$\pm$0.7 & 48.9$\pm$0.2 & 152.0$\pm$8.1\\
	   &  & 40-50 & 73$\pm$10  & 45.8$\pm$0.8 & 56.9$\pm$0.3 & 0.63$\pm$0.00 & 33.0$\pm$0.8 &  37.5$\pm$0.2 & 94.5$\pm$5.1\\
	   &  & 50-60 & 44$\pm$9   & 32.9$\pm$0.8 & 45.9$\pm$0.2 & 0.56$\pm$0.00 & 21.9$\pm$0.8 & 28.9$\pm$0.1 & 55.8$\pm$3.1\\ 
	   &  & 60-70 & 26$\pm$7   & 21.8$\pm$0.8 & 36.4$\pm$0.2 & 0.47$\pm$0.01 & 12.8$\pm$0.7 & 22.6$\pm$0.1 & 31.3$\pm$1.8\\
	   &  & 70-80 & 14$\pm$6   & 12.1$\pm$0.7 & 26.8$\pm$0.2 & 0.37$\pm$0.01 & 5.4$\pm$0.6 & 15.5$\pm$0.1 & 16.0$\pm$0.9\\
	   \hline	   
     &    &  0-5  & 338$\pm$2 & 146.6$\pm$0.7 & 147.5$\pm$0.7 & 0.89$\pm$0.00 & 126.9$\pm$0.6 & 125.0$\pm$0.6 & 683.4$\pm$40.0\\
       &  & 5-10 & 289$\pm$6 & 127.2$\pm$0.7 & 129.9$\pm$0.6 & 0.85$\pm$0.00 & 107.6$\pm$0.7 & 105.1$\pm$0.5 & 556.5$\pm$32.3\\
	   &  & 10-20 & 225$\pm$9 & 104.0$\pm$0.7 & 109.4$\pm$0.5 & 0.81$\pm$0.00 & 85.6$\pm$0.7 & 84.5$\pm$0.4 & 421.7$\pm$24.7\\
  &  & 20-30 & 158$\pm$10 & 80.2$\pm$0.8 & 87.7$\pm$0.4 & 0.76$\pm$0.00 & 63.6$\pm$0.7 & 64.2$\pm$0.3 & 284.3$\pm$16.6\\
Au-Au	   & 19.6 & 30-40 & 108$\pm$11 & 61.4$\pm$0.8 & 70.7$\pm$0.3 & 0.70$\pm$0.00 & 46.6$\pm$0.7 &  49.0$\pm$0.2 & 187.9$\pm$11.1\\
	   &  & 40-50 & 71$\pm$10 & 46.0$\pm$0.8 & 57.1$\pm$0.3 & 0.64$\pm$0.00 & 33.0$\pm$0.8 &  37.5$\pm$0.2 & 117.3$\pm$7.0\\
	   &  & 50-60 & 44$\pm$9  & 32.9$\pm$0.8 & 45.9$\pm$0.2 & 0.56$\pm$0.00 & 21.8$\pm$0.8 & 28.8$\pm$0.1 & 70.3$\pm$4.2\\
	   &  & 60-70 & 25$\pm$7  & 21.9$\pm$0.8 & 36.6$\pm$0.2 & 0.47$\pm$0.01 & 12.9$\pm$0.7 & 22.7$\pm$0.1 & 38.1$\pm$2.3\\
	   &  & 70-80 & 14$\pm$5  & 12.1$\pm$0.5 & 26.7$\pm$0.2 & 0.37$\pm$0.01 & 5.4$\pm$0.6 & 15.5$\pm$0.1 & 19.8$\pm$1.3\\
	   \hline	   
      &    &  0-5  & 343$\pm$2 & 147.2$\pm$0.7 & 148.3$\pm$0.7 & 0.89$\pm$0.00 & 127.6$\pm$0.6 & 125.8$\pm$0.6 & 727.0$\pm$42.2\\
       &  &  5-10 & 299$\pm$6 & 127.8$\pm$0.7 & 130.9$\pm$0.6 & 0.85$\pm$0.00 & 108.1$\pm$0.7 & 105.9$\pm$0.5 & 605.7$\pm$35.2\\
	   &  & 10-20 & 234$\pm$9 & 104.6$\pm$0.7 & 110.2$\pm$0.5 & 0.81$\pm$0.00 & 85.9$\pm$0.7 & 85.1$\pm$0.4 & 457.6$\pm$26.6\\
	   &  & 20-30 & 166$\pm$11 & 80.7$\pm$0.8 & 88.4$\pm$0.4 & 0.76$\pm$0.00 & 63.8$\pm$0.7 & 64.6$\pm$0.3 & 309.6$\pm$18.1\\
Au-Au  & 27 & 30-40 & 114$\pm$11 & 61.9$\pm$0.8 & 71.3$\pm$0.3 & 0.71$\pm$0.00 & 46.8$\pm$0.7 & 49.3$\pm$0.2 & 203.6$\pm$11.9\\
	   &  & 40-50 & 75$\pm$10 & 46.1$\pm$0.8 & 57.4$\pm$0.3 & 0.64$\pm$0.00 & 33.1$\pm$0.8 & 37.6$\pm$0.2 & 128.2$\pm$7.6\\
	   &  & 50-60 & 47$\pm$9 & 32.9$\pm$0.8 & 46.1$\pm$0.2 & 0.56$\pm$0.00 & 21.7$\pm$0.8 & 28.8$\pm$0.1 & 76.0$\pm$4.5\\
	   &  & 60-70 & 27$\pm$8 & 21.9$\pm$0.8 & 36.6$\pm$0.2 & 0.47$\pm$0.01 & 12.8$\pm$0.7 & 22.5$\pm$0.1 & 41.9$\pm$2.3\\
	   &  & 70-80 & 14$\pm$6 & 12.8$\pm$0.7 & 27.5$\pm$0.2 & 0.38$\pm$0.01 & 5.9$\pm$0.6 & 16.3$\pm$0.1 & 20.4$\pm$1.3\\
	   \hline 
       \hline
      &    &  0-5  & 342$\pm$2 & 147.9$\pm$0.7 & 149.2$\pm$0.7 & 0.89$\pm$0.00 & 128.4$\pm$0.6 & 126.7$\pm$0.6 & 756.2$\pm$44.1\\
       &  &  5-10 & 294$\pm$6 & 128.4$\pm$0.7 & 131.7$\pm$0.6 & 0.86$\pm$0.00 &  108.6$\pm$0.7 & 106.6$\pm$0.5 & 633.0$\pm$36.8\\
	   &  & 10-20 & 230$\pm$9 & 104.9$\pm$0.7 & 110.8$\pm$0.5 & 0.81$\pm$0.00 & 86.2$\pm$0.7 & 85.6$\pm$0.4 & 482.2$\pm$28.2\\
 &  & 20-30 & 162$\pm$10 & 81.0$\pm$0.8 & 89.0$\pm$0.4 & 0.76$\pm$0.00 & 64.0$\pm$0.7 & 65.0$\pm$0.3 & 328.2$\pm$19.1\\
Au-Au	   & 39 & 30-40 & 111$\pm$11 & 62.2$\pm$0.8 & 71.9$\pm$0.3 & 0.71$\pm$0.00 & 47.1$\pm$0.7 & 49.7$\pm$0.2 & 215.6$\pm$12.5\\
	   &  & 40-50 & 74$\pm$10 & 46.6$\pm$0.8 & 58.0$\pm$0.3 & 0.65$\pm$0.00 & 33.3$\pm$0.8 & 37.9$\pm$0.2 & 136.3$\pm$7.9\\
	   &  & 50-60 & 46$\pm$9 & 33.4$\pm$0.8 & 46.6$\pm$0.2 & 0.57$\pm$0.00 & 22.1$\pm$0.8 & 29.0$\pm$0.1 & 82.9$\pm$4.8\\
	   &  & 60-70 & 26$\pm$7 & 22.1$\pm$0.8 & 36.8$\pm$0.2 & 0.48$\pm$0.01 & 12.9$\pm$0.7 & 22.5$\pm$0.1 & 44.9$\pm$2.6\\
	   &  & 70-80 & 14$\pm$5 & 12.8$\pm$0.7 & 27.5$\pm$0.2 & 0.38$\pm$0.01 & 5.8$\pm$0.6 & 16.1$\pm$0.1 & 23.5$\pm$1.5\\
	   \hline
	   \hline
     &     &  0-5  & 346.5$\pm$2.8  & 148.9$\pm$0.7 & 150.1$\pm$0.7 & 0.90$\pm$0.00 & 129.1$\pm$0.6 & 127.6$\pm$0.6 & 952.8$\pm$37.7\\
       &  &  5-10 & 293.9$\pm$4.2  & 129.3$\pm$0.7 & 132.8$\pm$0.6 & 0.86$\pm$0.00 & 109.1$\pm$0.7 & 107.4$\pm$0.5 & 780.9$\pm$28.7\\
	   &  & 10-20 & 229.8$\pm$4.6  & 105.7$\pm$0.8 & 111.9$\pm$0.5 & 0.82$\pm$0.00 & 86.4$\pm$0.7 & 86.3$\pm$0.4 & 588.1$\pm$24.0\\
 &  & 20-30 & 164.1$\pm$5.4  & 81.5$\pm$0.8 & 89.9$\pm$0.4 & 0.77$\pm$0.00 & 64.1$\pm$0.7 & 65.4$\pm$0.3 & 406.4$\pm$15.3\\
Au-Au	   & 62.4 & 30-40 & 114.3$\pm$5.1  & 62.6$\pm$0.8 & 72.6$\pm$0.3 & 0.72$\pm$0.00 & 47.1$\pm$0.2 &  49.9$\pm$0.2 & 270.3$\pm$11.2\\
	   &  & 40-50 & 76.3$\pm$5.2   & 47.0$\pm$0.8 & 58.7$\pm$0.3 & 0.65$\pm$0.00 & 33.4$\pm$0.7 & 38.1$\pm$0.1 & 174.4$\pm$7.6\\
	   &  & 50-60 & 47.9$\pm$4.7   & 33.7$\pm$0.8 & 47.1$\pm$0.2 & 0.58$\pm$0.00 & 22.2$\pm$0.8 & 29.1$\pm$0.1 & 105.5$\pm$5.1\\
	   &  & 60-70 & 27.8$\pm$3.7   & 22.3$\pm$0.8 & 37.1$\pm$0.2 & 0.48$\pm$0.01 & 12.9$\pm$0.8 & 22.4$\pm$0.1 & 58.2$\pm$2.8\\
	   &  & 70-80 & 15.3$\pm$2.4 & 12.9$\pm$0.7 & 27.5$\pm$0.2 & 0.38$\pm$0.01 & 5.8$\pm$0.7 & 15.9$\pm$0.1 & 28.4$\pm$1.3\\
	   \hline	   
     &     &  0-6  & 344.3$\pm$3.1 & 148.4$\pm$0.7 & 150.3$\pm$0.7 & 0.90$\pm$0.00 & 128.6$\pm$0.7 & 127.4$\pm$0.6 & 1140.8$\pm$43.9\\
       &  &  6-11 & 289.0$\pm$5.4 & 127.3$\pm$0.7 & 131.8$\pm$0.6 & 0.86$\pm$0.00 & 106.8$\pm$0.7 & 106.0$\pm$0.5 & 920.6$\pm$35.1\\
	   &  & 11-18 & 237.8$\pm$6.8 & 108.7$\pm$0.7 & 115.3$\pm$0.5 & 0.83$\pm$0.00 & 88.9$\pm$0.7 & 89.1$\pm$0.4 & 751.0$\pm$34.9\\
	   &  & 18-26 & 187.7$\pm$7.5 & 89.5$\pm$0.8 & 97.8$\pm$0.4 & 0.79$\pm$0.00 & 71.0$\pm$0.7 & 72.4$\pm$0.3  & 569.3$\pm$24.2\\
Au-Au. & 130 & 26-34 & 141.9$\pm$8.4 & 72.6$\pm$0.8 & 82.3$\pm$0.4 & 0.75$\pm$0.00 & 55.6$\pm$0.7 & 58.2$\pm$0.3 & 371.3$\pm$17.6\\
	   &  & 34-45 & 100.9$\pm$8.4 & 56.0$\pm$0.8 & 67.2$\pm$0.3 & 0.70$\pm$0.00 & 40.9$\pm$0.8 & 44.9$\pm$0.2 & 246.0$\pm$11.0\\
	   &  & 45-58 & 61.0$\pm$7.8  & 38.7$\pm$0.8 & 51.9$\pm$0.3 & 0.61$\pm$0.00 & 26.2$\pm$0.8 & 32.4$\pm$0.2 & 131.6$\pm$7.6\\
	   &  & 58-85 & 22.6$\pm$5.0 & 16.3$\pm$0.8 & 31.5$\pm$0.2 & 0.43$\pm$0.01 & 8.2$\pm$0.8 & 17.7$\pm$0.1 & 39.9$\pm$4.3\\
	   \hline	   
    &      &  0-5  & 350.6$\pm$2.4 & 150.9$\pm$0.7 & 152.4$\pm$0.7 & 0.92$\pm$0.00  & 131.2$\pm$0.7 &  130.3$\pm$0.6 & 1335.8$\pm$57.1  \\
       &  &  5-10 & 298.6$\pm$4.1 & 131.6$\pm$0.7 & 136.0$\pm$0.6 & 0.88$\pm$0.00  & 110.9$\pm$0.7 &  110.4$\pm$0.5 & 1068.8$\pm$45.2 \\
	   &  & 10-20 & 234.3$\pm$4.6 & 108.1$\pm$0.8 & 115.3$\pm$0.5 & 0.83$\pm$0.00  & 88.1$\pm$0.7  &  89.1$\pm$0.4  & 797.2$\pm$33.5 \\
  &  & 20-30 & 167.6$\pm$5.4 & 83.6$\pm$0.8 &  93.0$\pm$0.4  & 0.79$\pm$0.00  & 65.4$\pm$0.7  &  67.8$\pm$0.3  & 553.6$\pm$22.2 \\
Au-Au	   & 200 & 30-40 & 117.1$\pm$5.2 & 64.6$\pm$0.8&  75.5$\pm$0.4  & 0.73$\pm$0.00  & 48.3$\pm$0.8  &  51.9$\pm$0.2  & 365.2$\pm$15.1 \\
	   &  & 40-50 & 78.3$\pm$5.3  & 48.3$\pm$0.8 &  60.8$\pm$0.3  & 0.68$\pm$0.00  & 34.1$\pm$0.8  &  39.2$\pm$0.2  & 238.9$\pm$9.9\\
	   &  & 50-60 & 49.3$\pm$4.7  & 34.9$\pm$0.8 &  48.9$\pm$0.2  & 0.59$\pm$0.00  & 22.8$\pm$0.8  &  29.7$\pm$0.1  & 147.8$\pm$5.9\\
	   &  & 60-70 & 28.8$\pm$3.7  & 23.8$\pm$0.8 & 39.1$\pm$0.2& 0.51$\pm$0.01 				   & 13.9$\pm$0.8& 22.7$\pm$0.1 & 85.2$\pm$3.5\\
	   &  & 70-80 & 15.7$\pm$2.6  & 13.2$\pm$0.8 & 28.2$\pm$0.2& 0.40$\pm$0.00                  & 6.2$\pm$0.7& 15.2$\pm$0.1 & 42.6$\pm$1.8\\
\hline
\hline
   &    &  0-5  & 382.5$\pm$3.1    & 166.9$\pm$0.7 & 170.7$\pm$0.7& 0.94$\pm$0.00  & 146.0$\pm$0.7 & 148.0$\pm$0.6 & 2837.0$\pm$144.0\\
       &  &  5-10 & 329.4$\pm$4.9 & 146.1$\pm$0.7 & 154.7$\pm$0.6 & 0.90$\pm$0.00  & 121.9$\pm$0.7& 126.5$\pm$0.6 & 2345.5$\pm$112.4\\
	   &  & 10-20 & 259.9$\pm$2.9 & 119.8$\pm$0.8 & 132.4$\pm$0.6& 0.86$\pm$0.00  & 96.3$\pm$0.7 & 102.7$\pm$0.6 & 1763.2$\pm$84.8\\
 &  & 20-30 & 185.4$\pm$3.9 & 92.9$\pm$0.8 & 107.5$\pm$0.5& 0.81$\pm$0.00 & 71.5$\pm$0.8 & 78.4$\pm$0.3 & 1195.8$\pm$54.2\\
Pb-Pb	   & 2760 & 30-40 & 128.1$\pm$3.3 & 71.4$\pm$0.8 & 87.2$\pm$0.4& 0.76$\pm$0.00 & 52.4$\pm$0.8 & 59.7$\pm$0.2 & 784.8$\pm$35.9\\
	   &  & 40-50 & 84.2$\pm$2.6  & 53.7$\pm$0.8 & 70.3$\pm$0.3& 0.70$\pm$0.00  & 37.2$\pm$0.8& 44.8$\pm$0.2 & 482.7$\pm$21.4\\
	   &  & 50-60 & 52.1$\pm$2.0  & 38.6$\pm$0.8 & 56.1$\pm$0.3& 0.63$\pm$0.00  & 24.7$\pm$0.9& 33.1$\pm$ 0.1 & 274.8$\pm$12.5\\
	   &  & 60-70 & 29.5$\pm$1.3  & 25.7$\pm$0.8 & 43.6$\pm$0.2& 0.54$\pm$0.00  & 14.6$\pm$0.9& 23.8$\pm$0.1 & 141.8$\pm$5.4\\
	   &  & 70-80 & 14.9$\pm$0.6 & 14.2$\pm$0.8 & 30.8$\pm$0.2 & 0.43$\pm$0.00  & 6.4$\pm$0.7& 15.1$\pm$0.1 & 67.2$\pm$3.0\\
	  \hline
   &    &  0-5 & 385$\pm$2 & 170.2$\pm$0.7  & 174.2$\pm$0.7 & 0.94$\pm$0.00 & 149.0$\pm$0.7 & 151.5$\pm$0.6 & 3320.6$\pm$131.4\\
       &  &  5-10 & 333$\pm$4     & 149.2$\pm$0.7  & 158.5$\pm$0.6 & 0.90$\pm$0.00 & 124.4$\pm$0.7 & 129.9$\pm$0.5 & 2698.7$\pm$117.2\\
	   &  & 10-20 & 263$\pm$4     & 122.4$\pm$0.8  & 135.8$\pm$0.6 & 0.86$\pm$0.00 & 98.1$\pm$0.7 & 105.6$\pm$0.4 & 2042.5$\pm$84.7\\
 &  & 20-30 & 188$\pm$3     & 94.9$\pm$0.8  & 110.5$\pm$0.5  & 0.82$\pm$0.00 & 72.9$\pm$0.7 & 80.8$\pm$0.3 & 1401.4$\pm$62.9\\
Pb-Pb	   & 5020 & 30-40 & 131$\pm$2     & 73.4$\pm$0.8  & 90.0$\pm$0.4 & 0.77$\pm$0.00 & 53.8$\pm$0.8 & 61.8$\pm$0.3 & 931.0$\pm$44.5\\
	   &  & 40-50 & 86.3$\pm$1.7  & 55.7$\pm$0.8  & 73.1$\pm$0.3 & 0.71$\pm$0.00 & 38.6$\pm$0.8 & 46.9$\pm$0.2 & 588.6$\pm$27.8\\
	   &  & 50-60 & 53.6$\pm$1.2  & 40.7$\pm$0.8  & 58.7$\pm$0.3 & 0.63$\pm$0.00   & 26.3$\pm$0.8 & 34.9$\pm$0.1 & 346.9$\pm$26.1\\
	   &  & 60-70 & 30.0$\pm$0.8  & 27.9$\pm$0.8 & 45.9$\pm$0.2 & 0.54$\pm$0.01   & 16.2$\pm$0.9 & 25.5$\pm$0.1 & 186.1$\pm$26.0\\
	   &  & 70-80 & 15.6$\pm$0.5  & 16.6$\pm$0.7 & 33.0$\pm$0.2 & 0.43$\pm$0.01  & 7.7$\pm$0.7 & 17.0$\pm$0.1 & 93.5$\pm$27.4\\
	  \hline       	            
\caption[Tabeeel.]{
For the colliding systems Au-Au at $\sqrt{s_{NN}}$=7.7, 11.5, 19.6, 27 and 39 GeV from BES \cite{STAR2} and Au-Au at $\sqrt{s_{NN}}$=62.4, 130 and 200~GeV \cite{STAR1} studied
at RHIC by the STAR Collaboration and Pb-Pb at $\sqrt{s_{NN}}$=2.76 and 5.02~TeV  
investigated by the ALICE Collaboration at LHC \cite{ALICE1, ALICE21}, are shown:
the colliding system, the collision energy, the centrality, the average number of participant nucleons in the collision ($\langle N_{part} \rangle$), the overlapping areas corresponding to the wounded nucleons ($S_{\perp}^{geom}$, $S_{\perp}^{var}$), estimated by the two recipes explained in the text, the percentage of the wounded nucleons undergoing more than a single collision ($f_{core}$),  the 
corresponding areas of the wounded nucleons undergoing more than a single collision
($(S_{\perp}^{geom})^{core}$, $(S_{\perp}^{var})^{core}$)
 and the hadron density 
($dN/dy$).}
\label{tab:tabel}
\end{longtable*}

\begin{figure}[th!]
\includegraphics[width=1.00\linewidth]{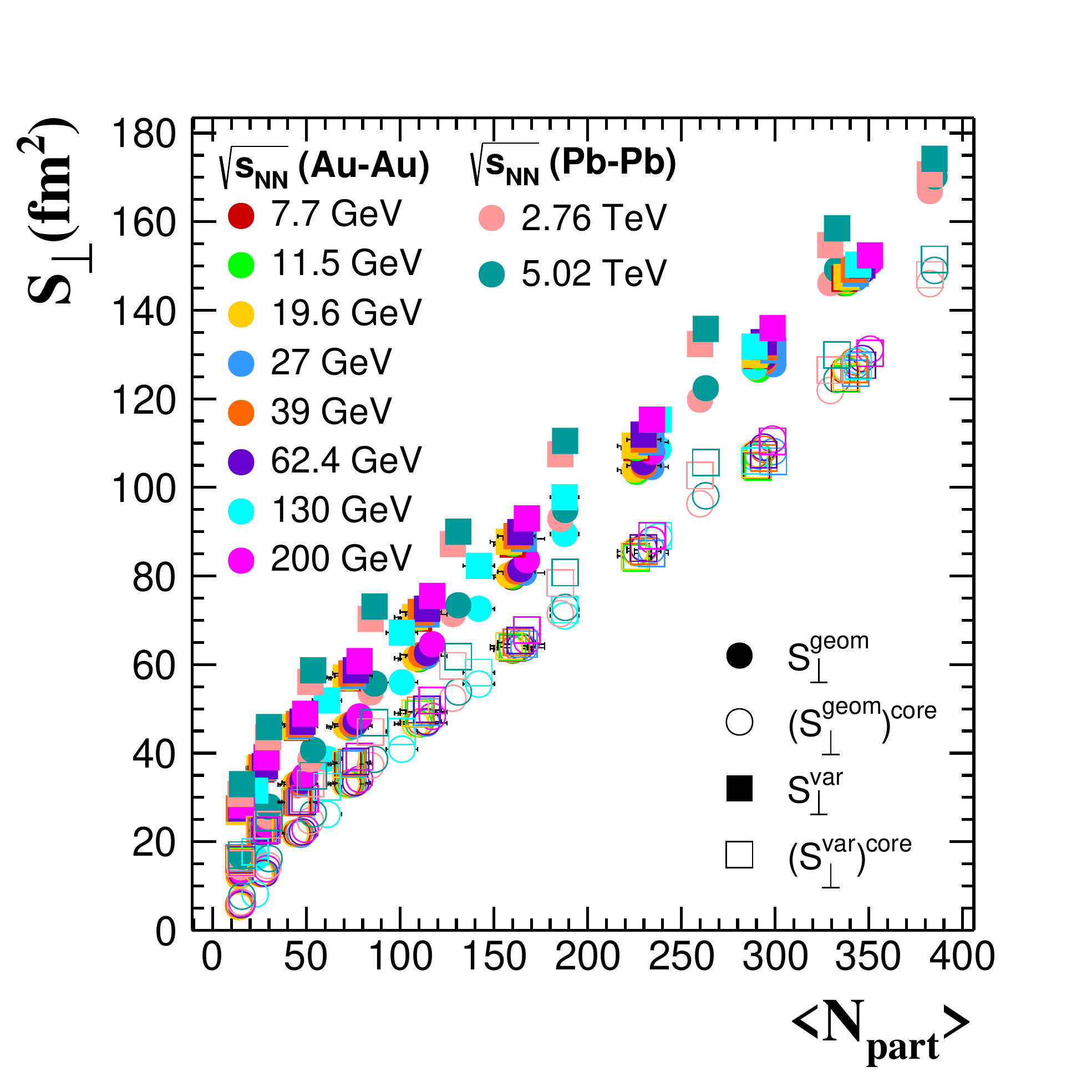}
\caption{Overlapping area of the colliding nuclei at different $\sqrt{s_{NN}}$ energies estimated within the GMC approach corresponding to all wounded nucleons $S_{\perp}^{geom}$ ($S_{\perp}^{var}$)
 - full dots (full squares)
and to the core contribution $(S_{\perp}^{geom})^{core}$
($(S_{\perp}^{var})^{core}$) - open dots (open squares) as a function of 
$\langle N_{part} \rangle$.}
\label{fig-1}
\end{figure}
\begin{figure}[th!]
\includegraphics[width=1.00\linewidth]{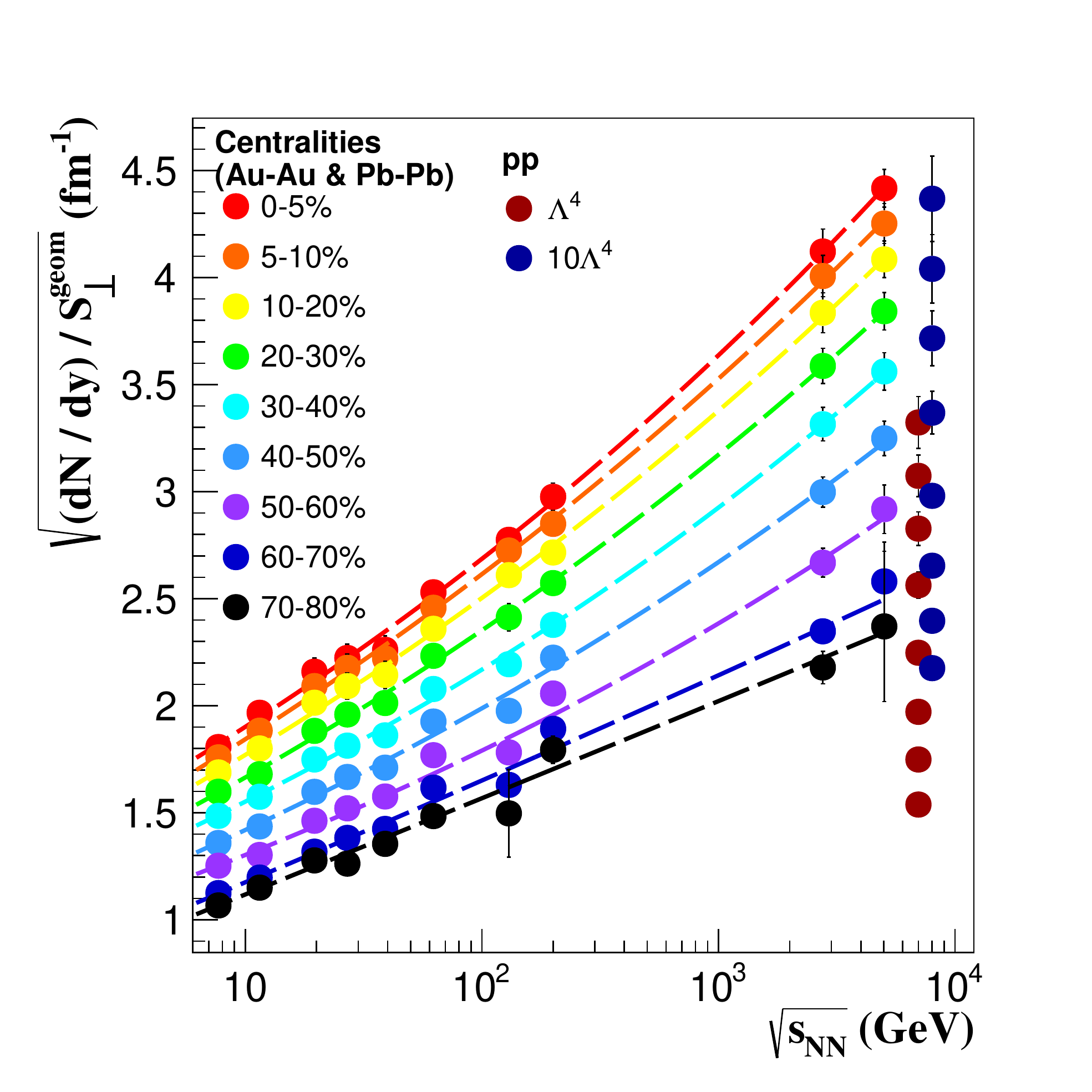}
\caption{$\sqrt{(dN/dy)/S_{\perp}^{geom}}$ as a function of $\sqrt{s_{NN}}$ for different centralities based on the values listed in Table I. The dashed lines 
represent the fit results using a power law function. Dark red and dark blue full dots correspond to
pp collision at \mbox{$\sqrt{s}$=7 TeV}, the values being estimated based on the IP-Glasma 
initial state model, using two values of the $\alpha$ parameter (see Chapter VII). 
For better clarity, the blue dots were artificially displaced in 
$\sqrt{s_{NN}}$.
}
\label{fig-2}
\end{figure}
\begin{figure}[th!]
\includegraphics[width=1.05\linewidth]{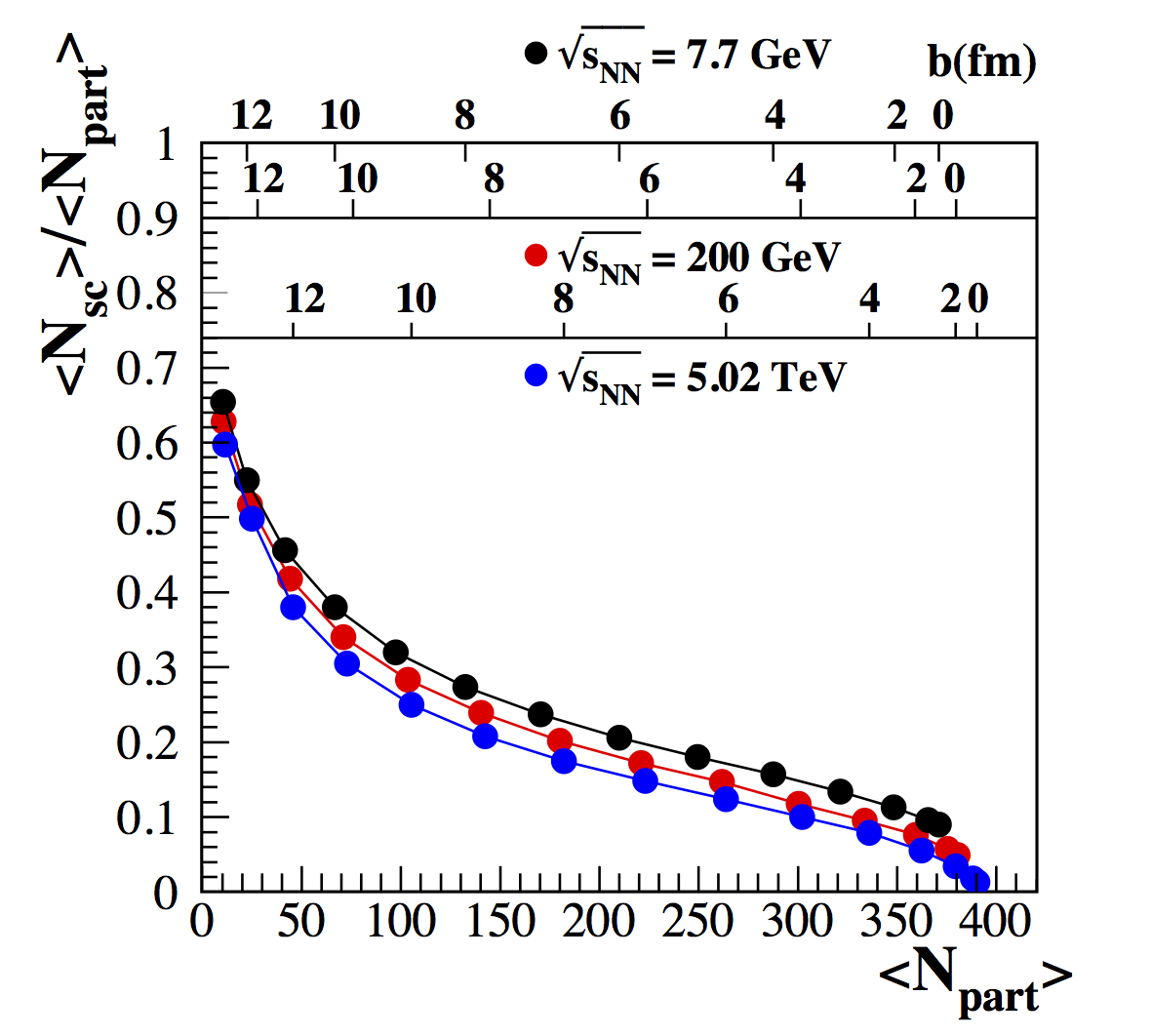}
\caption{The percentage of the nucleons suffering a single collision as a function of $\langle N_{part} \rangle$ and impact parameter for Au-Au collisions at $\sqrt{s_{NN}}$=7.7 and 200 GeV and Pb-Pb collisions at $\sqrt{s_{NN}}$=5.02 TeV.}
\label{fig-3}
\end{figure}
\noindent
The centrality dependent values were rescaled by the factor obtained dividing the
geometrical area to S in the case of the complete overlap of the nuclei (b=0 fm).
The $\langle N_{part} \rangle$ dependence of the   
overlap area 
of the colliding nuclei at different energies 
estimated within the GMC approach corresponding to all
wounded nucleons 
and to the core contribution 
are
presented in Fig.1.
$\sqrt{(dN/dy)/S_{\perp}^{geom}}$ as a function of $\sqrt{s_{NN}}$ 
for different 
centralities
is represented in Fig.2.
As an example, in Fig.3 
the percentage
of the nucleons suffering 
a single collision as a function of $\langle N_{part} \rangle$ and impact  
parameter is represented for the lowest and highest $\sqrt{s_{NN}}$ Au-Au collisions, 
i.e 7.7 and 200 GeV,  and for Pb-Pb at the highest LHC energy, $\sqrt{s_{NN}}$=5.02 TeV. As  expected, the $\langle N_{part}\rangle$ 
 dependence of the percentage of nucleons undergoing a
single collision 
is less dependent on $\sqrt{s_{NN}}$ than on the impact parameter.
\section{$\textbf{dN/dy}$ estimates}
   The total hadron density per unit of rapidity has been estimated based on the published identified charged hadrons
densities \cite{STAR2, STAR1, ALICE1, ALICE2} and hyperons densities
\cite{STAR3, STAR4, STAR5, STAR6, ALICE3, ALICE4, ALICE71}.
For $\sqrt{s_{NN}}$=19.6 and 27 GeV BES energies or some of the centralities, where the hyperon yields were not reported, the corresponding values were obtained by 
interpolation using the energy and centrality dependence fits. 
\begin{figure*}[t!]
\includegraphics[width=0.95\linewidth]{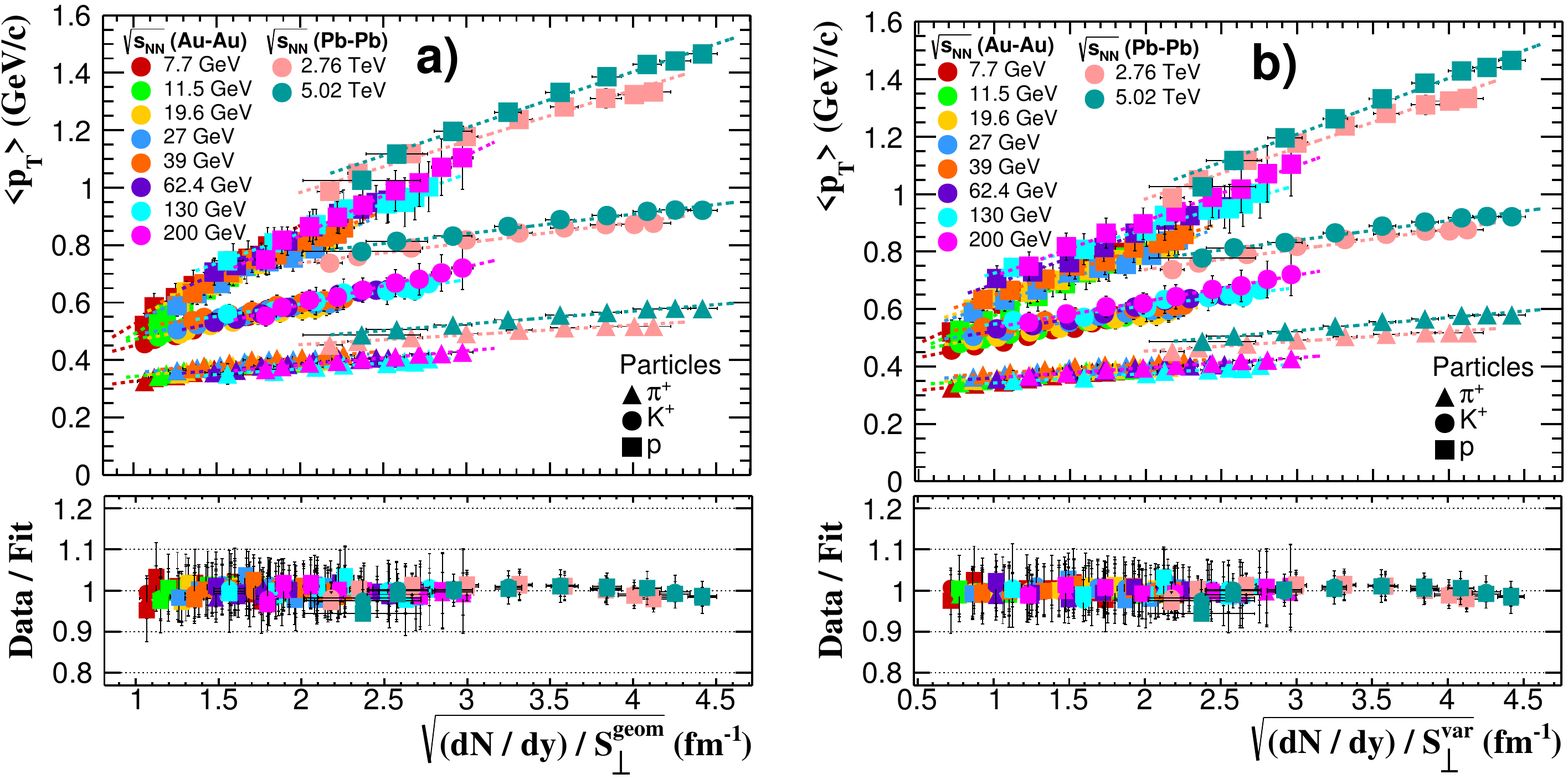}
\vspace{-1.5cm}
\caption{a) Top: $\langle p_T\rangle$
of pions, kaons and protons for all measured energies and centralities at RHIC and LHC reported by the 
STAR \cite{STAR1, STAR2}
and ALICE \cite{ALICE1, ALICE2} Collaborations, dashed lines representing
the results of the first order polynomial fit;
bottom: the ratio of the data points to the result of the linear fit  for each 
collision energy, as a function of 
$\sqrt{\frac{dN}{dy}/S_{\perp}^{geom}}$. 
b) Same as a) but as a function of $\sqrt{\frac{dN}{dy}/S_{\perp}^{var}}$.}
\label{fig-4}
\end{figure*}

As far as $\Omega^-$ and  $\bar{\Omega}^+$ 
yield values 
for BES were not reported and the extrapolation from higher energies down to 
BES energies shows a negligible contribution, they were not considered in the 
produced hadron density estimates.
Therefore, we used the following approximations: for the BES energies $\frac{dN}{dy}\simeq \frac{3}{2}\frac{dN}{dy}^{(\pi^+ + \pi^-)}+ 2\frac{dN}{dy}^{(K^++K^-, p+\bar{p}, \Xi^- +\bar{\Xi}^+)} + \frac{dN}{dy}^{( \Lambda + \bar{\Lambda})}$, 
from $\sqrt{s_{NN}}$=62.4 GeV to $\sqrt{s_{NN}}$=200 GeV
$\frac{dN}{dy}\simeq \frac{3}{2}\frac{dN}{dy}^{(\pi^+ + \pi^-)}+ 
2\frac{dN}{dy}^{(K^++K^-, p+\bar{p}, \Xi^- +\bar{\Xi}^+)} +
\frac{dN}{dy}^{(\Lambda + \bar{\Lambda}, \Omega^- + \bar{\Omega^+})}$ 
and for the LHC energies
$\frac{dN}{dy}\simeq \frac{3}{2}\frac{dN}{dy}^{(\pi^+ + \pi^-)}+ 
2\frac{dN}{dy}^{(p+\bar{p}, \Xi^- +\bar{\Xi}^+)} +
\frac{dN}{dy}^{(K^++K^-,K_S^0 + \bar{K_S^0}, \Lambda + \bar{\Lambda}, \Omega^- + \bar{\Omega}^+)}$.
The values are listed in the last column of Table I.
\begin{table*}
  \begin{tabular}{|c|c|c|c|c|c|c|}
    \hline
$\sqrt{s_{NN}}$ (GeV) &
      \multicolumn{3}{c|}{Slope} &
      \multicolumn{3}{c|}{Offset} \\
  \cline{2-7}
    & $\pi^+$ & $K^+$ & p & $\pi^+$ & $K^+$ & p \\
    \hline
    \hline
7.7 & 0.08 $\pm$ 0.02 & 0.15 $\pm$ 0.03 & 0.35 $\pm$ 0.07 & 0.25 $\pm$ 0.03 & 0.30 $\pm$ 0.04 & 0.18 $\pm$ 0.10 \\
\hline
11.5 & 0.05 $\pm$ 0.02 & 0.12 $\pm$ 0.03 & 0.33 $\pm$ 0.07 & 0.29 $\pm$ 0.04 & 0.35 $\pm$ 0.05 & 0.16 $\pm$ 0.10 \\
\hline
19.6 & 0.05 $\pm$ 0.02 & 0.11 $\pm$ 0.04 & 0.24 $\pm$ 0.05 & 0.30 $\pm$ 0.04 & 0.36 $\pm$ 0.06 & 0.31 $\pm$ 0.08 \\
\hline
27 & 0.05 $\pm$ 0.02 & 0.10 $\pm$ 0.03 & 0.25 $\pm$ 0.05 & 0.31 $\pm$ 0.04 & 0.39 $\pm$ 0.06 & 0.29 $\pm$ 0.08 \\
\hline
39 & 0.05 $\pm$ 0.02 & 0.08 $\pm$ 0.04 & 0.24 $\pm$ 0.05 & 0.31 $\pm$ 0.05 & 0.43 $\pm$ 0.07 & 0.33 $\pm$ 0.09 \\
\hline
\hline
62.4 & 0.05 $\pm$ 0.01 & 0.12 $\pm$ 0.02 & 0.26 $\pm$ 0.05 & 0.30 $\pm$ 0.03 & 0.35 $\pm$ 0.04 & 0.31 $\pm$ 0.09 \\
\hline
130 & 0.04 $\pm$ 0.02 & 0.09 $\pm$ 0.02 & 0.21 $\pm$ 0.05 & 0.29 $\pm$ 0.04 & 0.42 $\pm$ 0.05 & 0.43 $\pm$ 0.10 \\
\hline
200 & 0.05 $\pm$ 0.02 & 0.13 $\pm$ 0.04 & 0.28 $\pm$ 0.06 & 0.28 $\pm$ 0.04 & 0.33 $\pm$ 0.09 & 0.27 $\pm$ 0.13 \\
\hline
\hline
2760 & 0.03 $\pm$ 0.01 & 0.07 $\pm$ 0.01 & 0.18 $\pm$ 0.02 & 0.39 $\pm$ 0.03 & 0.60 $\pm$ 0.05 & 0.63 $\pm$ 0.05 \\
\hline
5020 & 0.05 $\pm$ 0.01 & 0.07 $\pm$ 0.01 & 0.19 $\pm$ 0.02 & 0.39 $\pm$ 0.04 & 0.64 $\pm$ 0.04 & 0.63 $\pm$ 0.06 \\
\hline
\end{tabular}
\caption{The parameters for the linear fit of the $\langle p_T\rangle$
dependence on $\sqrt{\frac{dN}{dy}/S_{\perp}^{geom}}$ for pions, kaons and protons 
corresponding to the energies mentioned in the first column.}
\label{tab:tabel.}
\end{table*}

\begin{table*}  
\begin{tabular}{|c|c|c|c|c|c|c|}    
\hline   
$\sqrt{s_{NN}}$ (GeV) &
      \multicolumn{3}{c|}{Slope} &
      \multicolumn{3}{c|}{Offset} \\
  \cline{2-7}
    & $\pi^+$ & $K^+$ & p & $\pi^+$ & $K^+$ & p \\  
\hline    \hline
7.7 & 0.05 $\pm$ 0.02 & 0.11 $\pm$ 0.02 & 0.24 $\pm$ 0.05 & 0.29 $\pm$ 0.02 & 0.38
$\pm$ 0.03 & 0.36 $\pm$ 0.06 \\\hline11.5 & 0.04 $\pm$ 0.02 & 0.09 $\pm$ 0.02 & 0.24
$\pm$ 0.04 & 0.32 $\pm$ 0.02 & 0.42 $\pm$ 0.03 & 0.34 $\pm$ 0.06 \\\hline19.6 & 0.04
$\pm$ 0.02 & 0.08 $\pm$ 0.02 & 0.17 $\pm$ 0.03 & 0.33 $\pm$ 0.03 & 0.42 $\pm$ 0.04 &
0.46 $\pm$ 0.05 \\\hline27 & 0.03 $\pm$ 0.02 & 0.07 $\pm$ 0.02 & 0.18 $\pm$ 0.03 &
0.34 $\pm$ 0.03 & 0.45 $\pm$ 0.04 & 0.44 $\pm$ 0.05 \\\hline39 & 0.03 $\pm$ 0.02 &
0.06 $\pm$ 0.03 & 0.17 $\pm$ 0.03 & 0.34 $\pm$ 0.03 & 0.48 $\pm$ 0.04 & 0.48 $\pm$
0.05 \\\hline\hline62.4 & 0.03 $\pm$ 0.01 & 0.09 $\pm$ 0.02 & 0.18 $\pm$ 0.03 & 0.33
$\pm$ 0.02 & 0.44 $\pm$ 0.03 & 0.51 $\pm$ 0.06 \\\hline130 & 0.03 $\pm$ 0.01 & 0.06
$\pm$ 0.02 & 0.15 $\pm$ 0.03 & 0.31 $\pm$ 0.03 & 0.48 $\pm$ 0.04 & 0.57 $\pm$ 0.06
\\\hline200 & 0.04 $\pm$ 0.01 & 0.09 $\pm$ 0.03 & 0.19 $\pm$ 0.04 & 0.33 $\pm$ 0.03
& 0.44 $\pm$ 0.05 & 0.52 $\pm$ 0.08 \\\hline\hline2760 & 0.03 $\pm$ 0.01 & 0.07
$\pm$ 0.01 & 0.18 $\pm$ 0.02 & 0.39 $\pm$ 0.03 & 0.60 $\pm$ 0.04 & 0.63 $\pm$ 0.05
\\\hline5020 & 0.05 $\pm$ 0.01 & 0.07 $\pm$ 0.01 & 0.19 $\pm$ 0.02 & 0.39 $\pm$ 0.04
& 0.64 $\pm$ 0.04 & 0.62 $\pm$ 0.06 \\\hline   
\end{tabular}
\caption{The parameters for the linear fit of the $\langle p_T\rangle$ dependence
on $\sqrt{\frac{dN}{dy}/S_{\perp}^{var}}$ for pions, kaons and protons 
corresponding to the energies mentioned in the first column.}
\label{tab:tabel.}
\end{table*}
\begin{table*}
  \begin{tabular}{|c|c|c|c|c|c|c|}
    \hline
$\sqrt{s_{NN}}$ (GeV) &
      \multicolumn{3}{c|}{Slope} &
      \multicolumn{3}{c|}{Offset} \\
  \cline{2-7}
    & $\pi^+$ & $K^+$ & p & $\pi^+$ & $K^+$ & p \\
    \hline
    \hline
   200 & 0.05 $\pm$ 0.02 & 0.13 $\pm$ 0.04 & 0.28 $\pm$ 0.06 & 0.28 $\pm$ 0.04 & 0.33 $\pm$ 0.09 & 0.27 $\pm$ 0.13\\
   \hline
   2760 & 0.04 $\pm$ 0.01 & 0.09 $\pm$ 0.02 & 0.20 $\pm$ 0.03 & 0.37 $\pm$ 0.04 & 0.56 $\pm$ 0.07 & 0.56 $\pm$ 0.08\\
   \hline
   5020 & 0.05 $\pm$ 0.02 & 0.08 $\pm$ 0.02 & 0.22 $\pm$ 0.03 & 0.37 $\pm$ 0.06 & 0.60 $\pm$ 0.07 & 0.54 $\pm$ 0.10\\
   \hline
\end{tabular}
\caption{The parameters of the linear fit of $\langle p_T\rangle$
as a function of $\sqrt{(\frac{dN}{dy})/S_{\perp}^{geom}}$ for pions, kaons and protons 
corresponding to $\sqrt{s_{NN}}$=200 GeV, 2.76 TeV and 5.02 TeV collision energies. The very last three points at $\sqrt{s_{NN}}$=2.76 and 5.02 TeV were not
included in the fit.
}
\end{table*}
\begin{table*}
  \begin{tabular}{|c|c|c|c|c|c|c|}
    \hline
$\sqrt{s_{NN}}$ (GeV) &
      \multicolumn{3}{c|}{Slope} &
      \multicolumn{3}{c|}{Offset} \\
  \cline{2-7}
    & $\pi^+$ & $K^+$ & p & $\pi^+$ & $K^+$ & p \\
    \hline
    \hline
   200 & 0.02 $\pm$ 0.03 & 0.09 $\pm$ 0.06 & 0.20 $\pm$ 0.11 & 0.36 $\pm$ 0.07 & 0.43 $\pm$ 0.15 & 0.50 $\pm$ 0.29\\
   \hline
   2760 & 0.03 $\pm$ 0.02 & 0.07 $\pm$ 0.03 & 0.17 $\pm$ 0.04 & 0.40 $\pm$ 0.06 & 0.58 $\pm$ 0.10 & 0.66 $\pm$ 0.14\\
   \hline
   5020 & 0.03 $\pm$ 0.03 & 0.06 $\pm$ 0.02 & 0.17 $\pm$ 0.04 & 0.41 $\pm$ 0.11 & 0.65 $\pm$ 0.08 & 0.73 $\pm$ 0.16\\
   \hline
\end{tabular}
\caption{The parameters of the linear fit of $\langle p_T\rangle$
as a function of $\sqrt{(\frac{dN}{dy})^{core}/S_{\perp}^{core}}$ for pions, kaons and protons 
corresponding to $\sqrt{s_{NN}}$=200 GeV, 2.76 TeV and 5.02 TeV collision energies.
For $\sqrt{s_{NN}}$=2.76 TeV and 5.02 TeV, the last three centralities, where a levelling off is
evidenced, were not included in the fit.}
\end{table*}
\section{$\sqrt{\frac{dN}{dy}/S_{\perp}}$ dependence of $\langle p_T\rangle$}

   As it was already mentioned in the Introduction, in the local parton-hadron duality approach \cite{LPHD}, 
$\langle p_T\rangle$/$\sqrt{\frac{dN}{dy}/S_{\perp}}$ is proportional with $\frac{1}{n\sqrt{n}}$ where n is the number of charged 
hadrons produced via gluon fragmentation \mbox{\cite{Lev1, Lap1}}. Therefore, neglecting other effects like collective 
hydrodynamic expansion and suppression,  $\langle p_T\rangle$/$\sqrt{\frac{dN}{dy}/S_{\perp}}$ is expected to 
decrease in central collisions and at higher energies. $\langle p_T\rangle$ for Au-Au collisions at 
$\sqrt{s_{NN}}$=7.7, 11.5, 19.6, 27, 39 GeV \cite{STAR2}; $\sqrt{s_{NN}}$=62.4, 130, 200 GeV \cite{STAR1} and 
Pb-Pb collisions at $\sqrt{s_{NN}}$=2.76, 5.02 TeV \cite{ALICE1, ALICE2} for positive pions, kaons and
protons are represented as 
a function of $\sqrt{\frac{dN}{dy}/S_{\perp}}$ in Fig.4a for 
$S_{\perp}^{geom}$ and in Fig.4b for $S_{\perp}^{var}$. The data
points corresponding to each collision energy were fitted with a first order
polynomial function. The trends in the two figures are rather similar and the fit quality, in terms of Data/Fit ratios, presented 
in the bottom plots of Fig.4 is equally good. 
The fit parameters are listed in Table II and Table III for 
$S_{\perp}=S_{\perp}^{geom}$ and $S_{\perp}=S_{\perp}^{var}$,
respectively and 
represented in Fig.5.

The slope value increases from pions to protons. Although the 
experimental error bars are rather large at the RHIC energies,
a systematic decrease of the slopes with the collision energy is 
evidenced for the $\langle p_T\rangle$ dependence on
$\sqrt{\frac{dN}{dy}/S_{\perp}^{geom}}$ - full symbols. This trend is enhanced going from pions to protons.
The offset values are rather similar at the RHIC energies and increase for all the three species at LHC energies.
Using $S_{\perp}^{var}$, Fig.4b, the extracted slopes,
represented in Fig.5a by open symbols show a marginal variation as a function of 
collision energy - dashed lines. The corresponding offsets, represented in Fig.5b by open symbols,
within the error bars, are the same for pions and kaons and are systematically 
larger for protons at RHIC energies compared with the ones corresponding to
$S_{\perp}^{geom}$. One should remark that at LHC energies, 
the results using $S_{\perp}^{geom}$ or $S_{\perp}^{var}$ are the same.   
\begin{figure}[]
\includegraphics[width=1.10\linewidth]{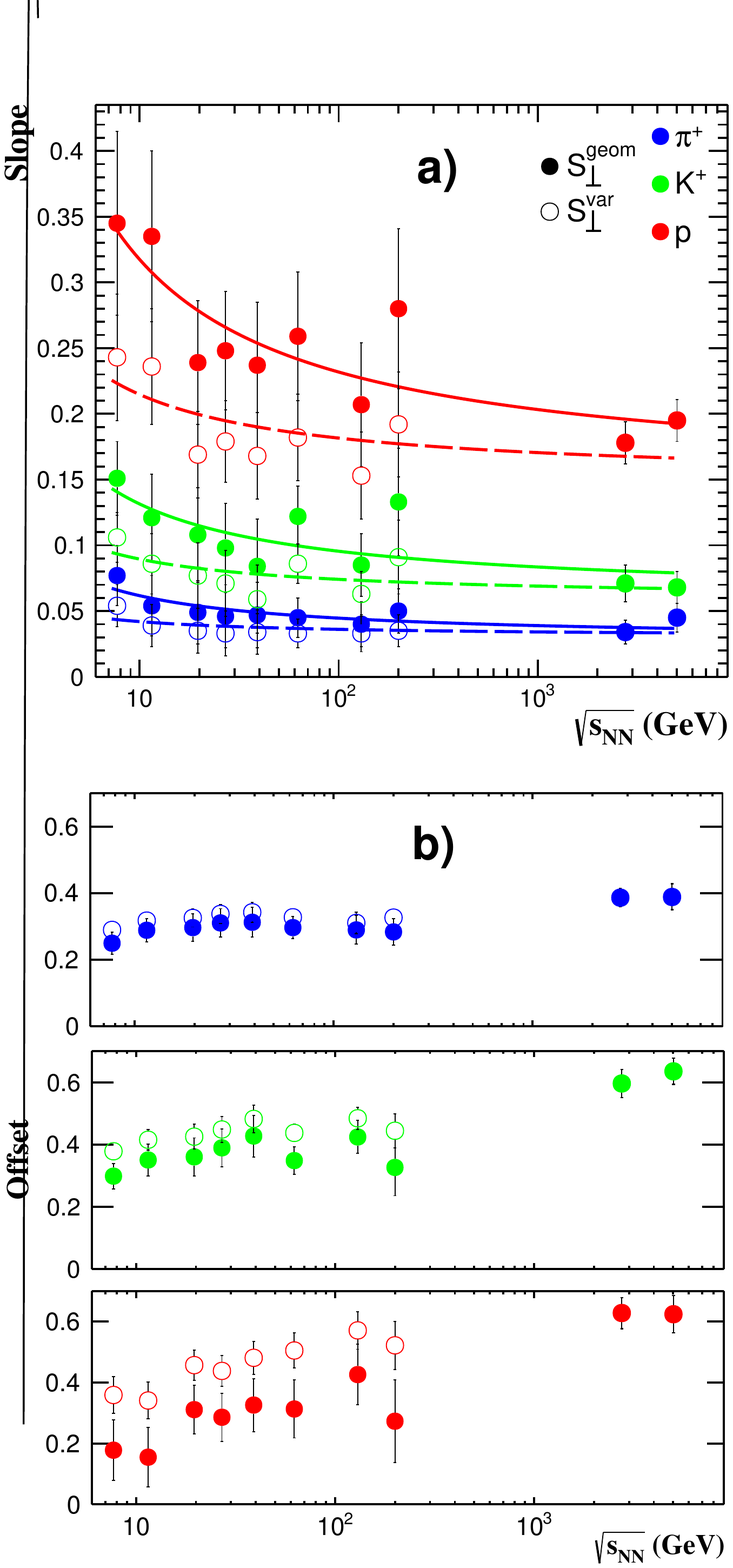}
\caption{a) The slopes of the $\langle p_T\rangle$ dependence on 
$\sqrt{\frac{dN}{dy}/S_{\perp}^{geom}}$ (full symbols) and on 
$\sqrt{\frac{dN}{dy}/S_{\perp}^{var}}$ (open symbols) for
$\pi^+$ (blue), $K^+$ (green) and $p$ (red) as a function of $\sqrt{s_{NN}}$. 
The fit results with the function
$a$+$b/(log\sqrt{s_{NN}})$ are drawn with full ($S_{\perp}^{geom}$) and dashed 
($S_{\perp}^{var}$) lines. b) The corresponding offsets.
}
\label{fig-5}
\end{figure}
\noindent
At the LHC energies, in the most central collisions, a saturation trend seems to develop. A natural question which comes is how much of the
observed trends is due to core-corona interplay \cite{Bec, Boz, Wer, Bec1, Aich, Boz1, Gem, Pet1} and how the
$\langle p_T\rangle$-$\sqrt{\frac{dN}{dy}/S_{\perp}}$ correlation for core looks like. Based on the recipe presented 
in \cite{Pet1}, we estimated the $\langle p_T\rangle^{core}$ for pions, kaons and protons for 
$\sqrt{s_{NN}}$=200 GeV, 2.76 TeV and 
5.02 TeV: 
\begin{multline}
\langle p_T\rangle _i^{cen}=\\
\frac{f_{core}\langle p_T\rangle_i^{core}M_i^{core}+(1-f_{core})\langle p_T\rangle_i^{ppMB}
M_i^{ppMB}}
{f_{core}M_i^{core}+(1-f_{core})
M_i^{ppMB}}
\end{multline}
$\langle p_T\rangle _i^{pp MB}$ for $\mathrm{\pi^+}$, $K^+$, p in pp minimum bias (MB) collisions at 
$\sqrt{s_{NN}}$=200 GeV were reported by the STAR 
Collaboration \cite{STAR1} and at $\sqrt{s_{NN}}$=2.76 TeV and 5.02 TeV were reported by the ALICE Collaboration \cite{ALICE7,ALICE72}. 
$(dN/dy)^{core}$ at the same energies were estimated using:  
\begin{equation}
\left( \frac{dN}{dy}\right) _i^{cen}=\langle N_{part}\rangle [(1-f_{core})M_i^{pp MB} + f_{core}M_i^{core}]
\end{equation}
\noindent
where $M_i^{ppMB}$=$\frac{1}{2}(dN/dy)_i^{ppMB}$ at the same energy and $M_i^{core}$ is the multiplicity per core
participant.  
$(dN/dy)_i^{ppMB}$ for $\mathrm{\pi^+}$, $K^+$, p were obtained based on the MB $p_T$ spectra reported in \cite{ALICE7,ALICE72}.

\begin{figure}[t!]
\includegraphics[width=1.1\linewidth]{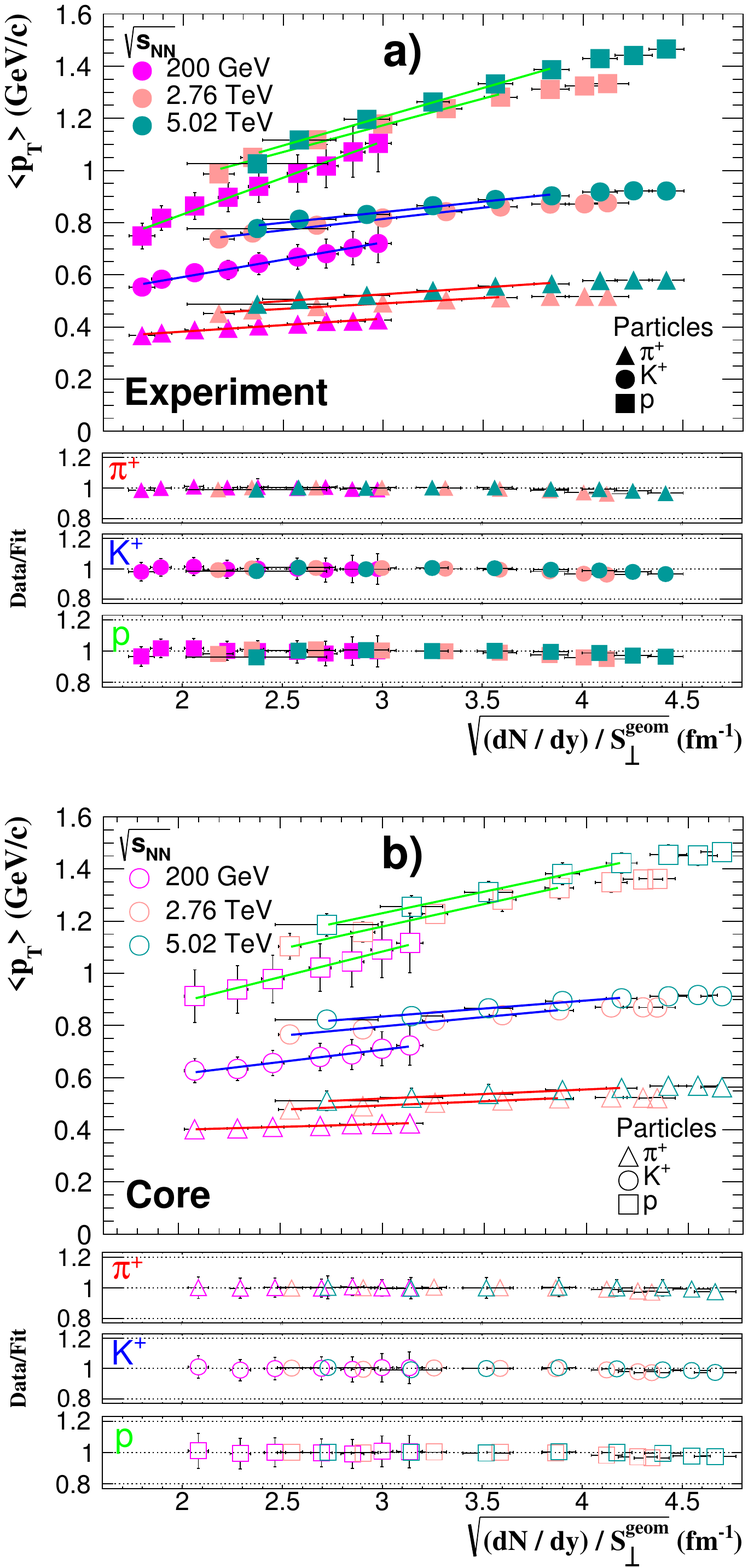}
\caption{$\langle p_T\rangle$ as a function of $\sqrt{\frac{dN}{dy}/S_{\perp}^{geom}}$ for identified charged hadrons for $\sqrt{s_{NN}}$=200 GeV, 2.76 TeV and 5.02 TeV.
The full lines represent the results of the first order polynomial fit.
a) Top: experimental results; Bottom: Data/Fit ratio; b) Top: estimated core contribution; Bottom: Data/Fit ratio.}
\label{fig-6}
\end{figure}
  
In Fig.6a $\langle p_T\rangle$ as a function of $\sqrt{\frac{dN}{dy}/S_{\perp}^{geom}}$
for pions, kaons and protons for $\sqrt{s_{NN}}$=200~GeV, 2.76~TeV and 5.02~TeV is represented.
The experimental points for each energy and each species were fitted with linear functions. As it was
already mentioned above, at $\sqrt{s_{NN}}$=2.76 TeV and 5.02 TeV the very last three points, corresponding to the most central 
collisions, systematically deviate
from a linear trend observed at lower centralities and therefore were excluded from the fit. 
The slopes and the offsets are presented in Table IV.
The fit quality 
can be followed in the bottom plot of Fig.6a where the ratios between the data 
points and fit results are represented.  
One can also observe that the last three points 
at $\sqrt{s_{NN}}$=2.76 TeV and 5.02 TeV, corresponding to the most central 
collisions, deviate from the general trend, the ratio $\langle p_T\rangle$/$\sqrt{\frac{dN}{dy}/S_{\perp}}$
is decreasing, as expected in Ref.\cite{Lev1}. 

For $\sqrt{s_{NN}}$=200 GeV, 2.76 TeV and 5.02 TeV we estimated    
$\langle p_T\rangle^{core}$ and $\sqrt{\frac{dN}{dy}^{core}/(S_{\perp}^{geom})^{core}}$, the results being presented in Fig.6b. 

The quality of the linear fit, represented in the bottom plot of Fig.6b is equally good as for the experimental data but the slope values presented in Table V are systematically 
smaller and the difference between the highest RHIC energy and the LHC energies is 
reduced. 
The saturation towards the
most central collisions at LHC energies does not change.   
\section{$\langle p_T\rangle$ particle mass dependence as a function of $\sqrt{\frac{dN}{dy}/S_{\perp}^{geom}}$}
\begin{figure}[t!]
\includegraphics[width=0.85\linewidth]{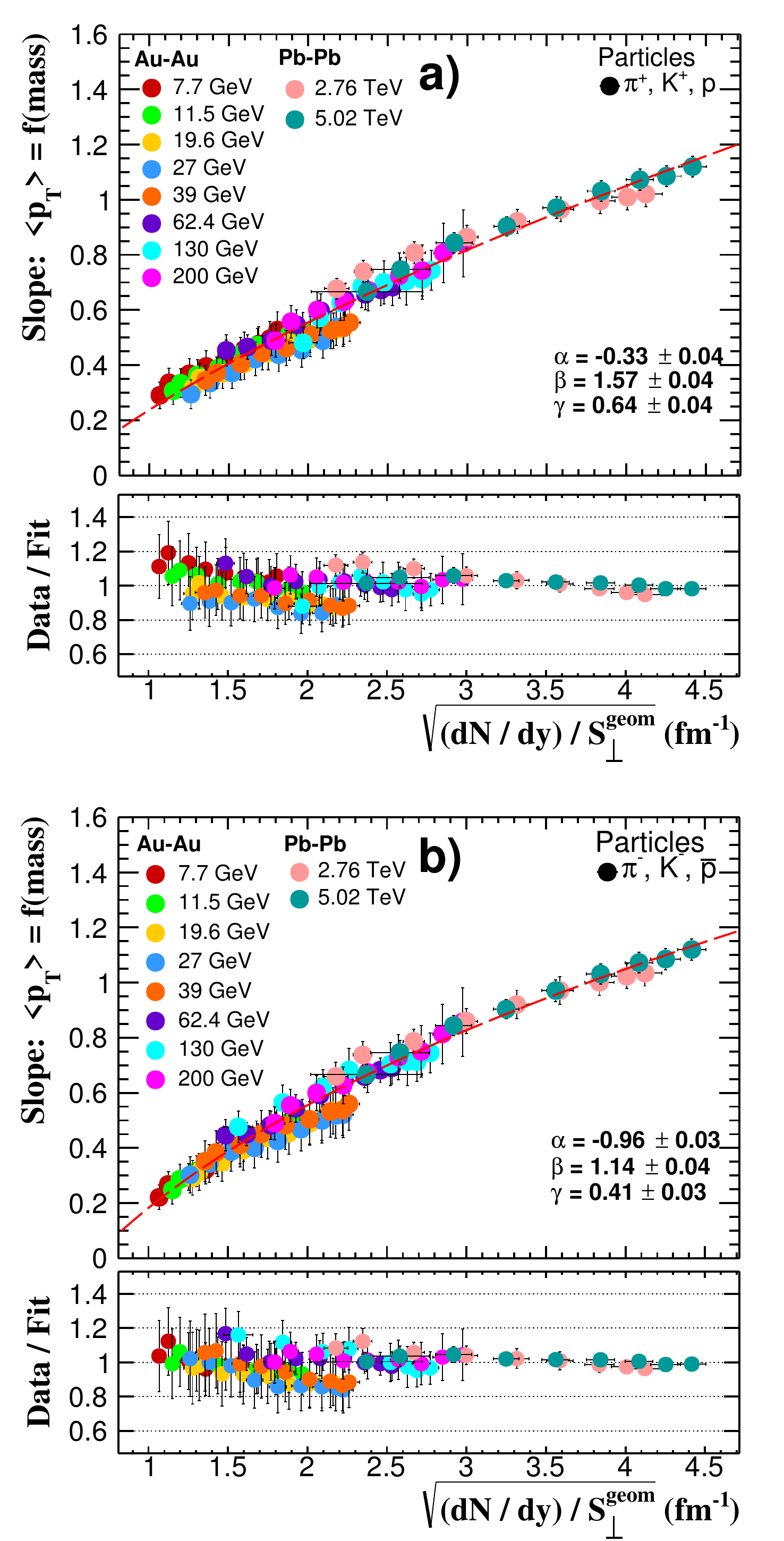}
\caption{The slopes from the linear fit of $\langle p_T\rangle$ versus particle mass as a function of $\sqrt{\frac{dN}{dy}/S_{\perp}^{geom}}$, for each centrality and energy,
for a) $\pi^+$, $K^+$ $p$ 
and 
b) $\pi^-$, $K^-$, $\bar{p}$. The continuous red line is the result of the fit with the function from Eq.4. The ratios Data/Fit are represented in the bottom 
plots of each of the two figures.
}
\label{fig-7}
\end{figure}   
   The $\langle p_T\rangle$ dependence on the mass of pions, kaons and protons at different collision centralities,
except for the most peripheral ones, is linear. Therefore,
linear fits of the $\langle p_T\rangle$ particle mass dependence, corresponding to each centrality 
and energy considered in the paper, were performed.
The extracted fit parameters as a function of $\sqrt{\frac{dN}{dy}/S_{\perp}^{geom}}$ 
are shown in Fig.7 (slope) and Fig.8 (offset).  
In Fig.7 the slopes are fitted with the following expression:
\begin{equation}
Slope_{\langle p_T\rangle = f(mass)}=\alpha + \beta\left(\sqrt{\frac{dN}{dy}/S_{\perp}^{geom}}\right)^{\gamma}
\end{equation}   
\begin{figure}[t!]
\includegraphics[width=0.9\linewidth]{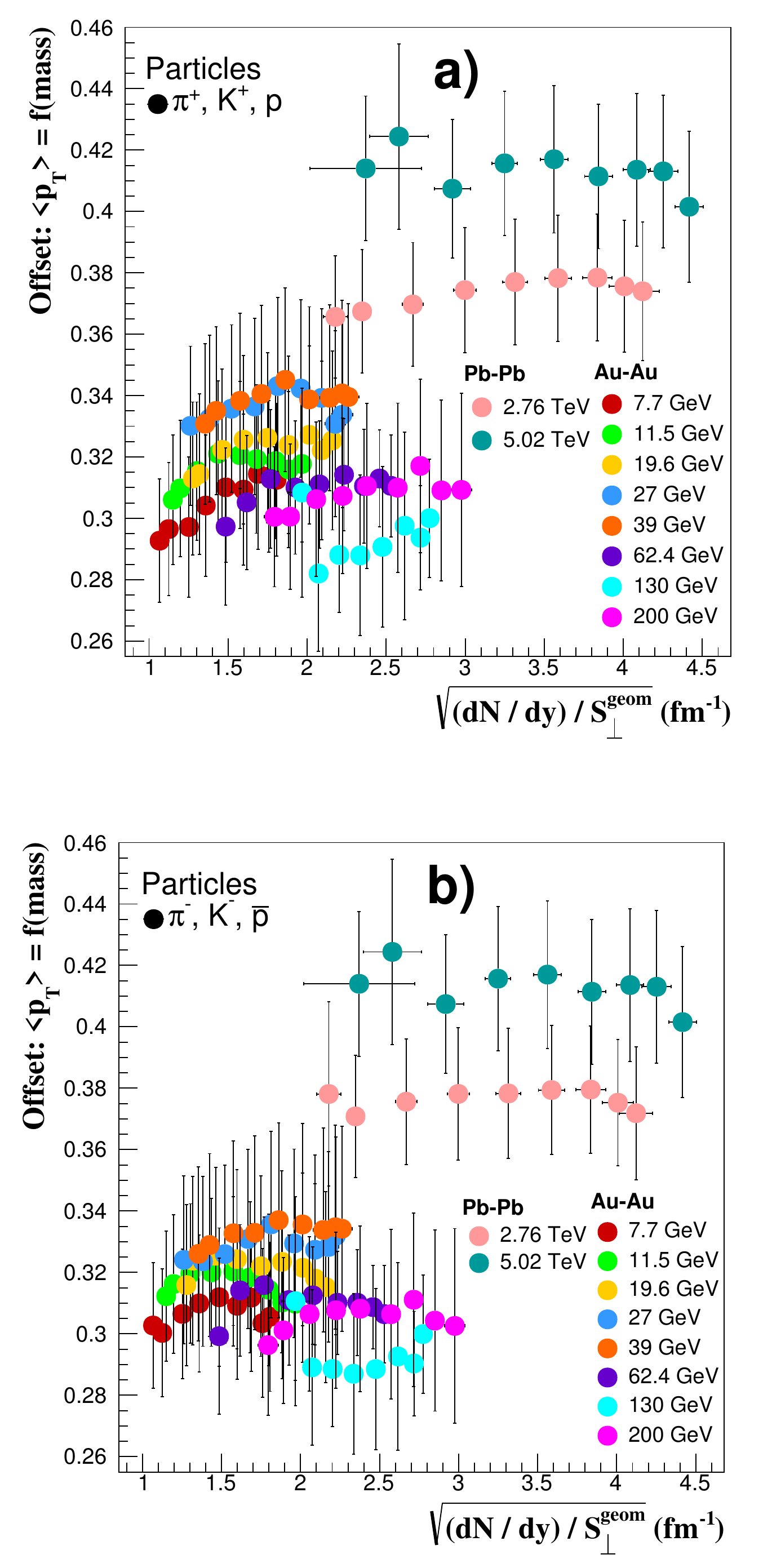}
\caption{The offsets from the linear fits of $\langle p_T\rangle$ versus particle mass dependence as a function of $\sqrt{\frac{dN}{dy}/S_{\perp}^{geom}}$;
a) particles, b) antiparticles.}
\label{fig-8}
\end{figure}
   The slopes for particles Fig.7a and antiparticles Fig.7b evidence a 
$\sqrt{\frac{dN}{dy}/S_{\perp}^{geom}}$ dependence which closely follows a trend given by Eq.4 (dashed lines). 
The values of the fit parameters are listed in Fig.7.
The fit quality is represented in the bottom 
plots of Fig.7a and Fig.7b in terms of Data/Fit. Besides the points corresponding to the most central collisions at $\sqrt{s_{NN}}$=19.6, 27 and 39~GeV which deviate from the fit by $\sim$10-15\%, the bulk of data nicely 
cluster around the fit curve, well within the error bars. 
In Fig.8, although 
the error bars are rather large, a systematic increase of the offsets as a function of $\sqrt{\frac{dN}{dy}/S_{\perp}^{geom}}$
is evidenced at BES energies and at $\sqrt{s_{NN}}$=62.4 GeV, reaching a plateau above 1.7~$fm^{-1}$. This trend is much reduced starting 
from $\sqrt{s_{NN}}$=130~GeV. Therefore, we considered only offsets above 1.7~$fm^{-1}$ and found their average values for different
$\sqrt{s_{NN}}$. The results are presented in Fig.9.
\begin{figure}[t!]
\includegraphics[width=0.95\linewidth]{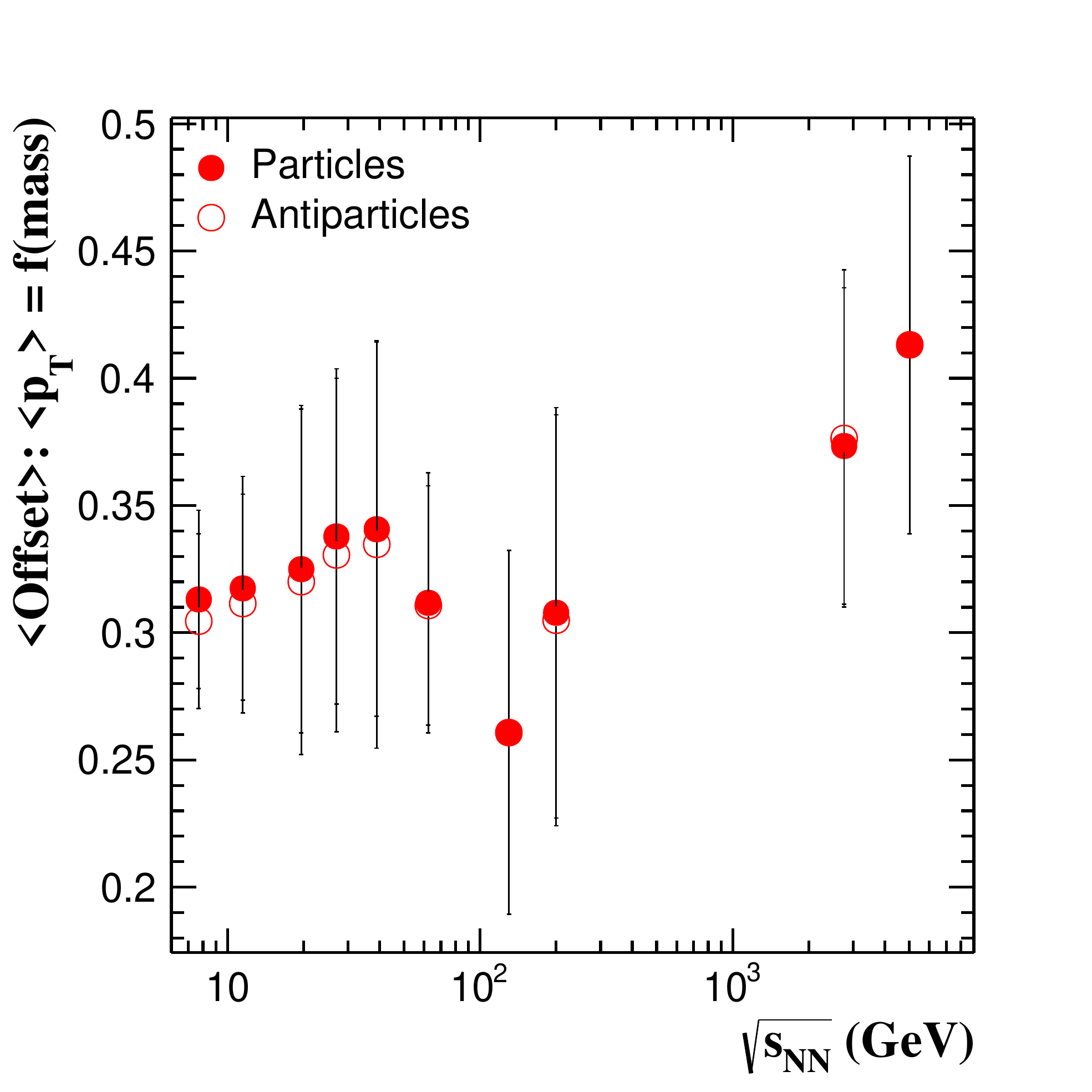}
\caption{The average values of the offsets of the $\langle p_T\rangle$ mass dependence for 
$\sqrt{\frac{dN}{dy}/S_{\perp}^{geom}}\geq$ 1.7 $fm^{-1}$ as a function of $\sqrt{s_{NN}}$.
Full symbols - particles; open symbols - antiparticles.}
\label{fig-9}
\end{figure}  

\section{$\sqrt{\frac{dN}{dy}/S_{\perp}^{geom}}$ dependence of Boltzmann-Gibbs Blast Wave fit parameters}

The $p_T$ spectra for identified charged hadrons were fitted \cite{STAR2, STAR1, ALICE1, ALICE2, ALICE22} using the BGBW expression inspired by hydrodynamic models \cite{Heinz}:
\begin{equation}
E\frac{d^3N}{dp^3}\sim
\int_{0}^Rm_TK_1(m_Tcosh\rho/T_{kin}^{fo})I_0(p_Tsinh\rho/T_{kin}^{fo})rdr
\end{equation}
where \mbox{$m_T=\sqrt{m^2+p_T^2}$; $\beta_T(r)=\beta_s(\frac{r}{R})^n$; $\rho=tanh^{-1}\beta_T$}.
$T_{kin}^{fo}$ is the kinetic freeze-out temperature and n defines the expansion profile.
A compilation of all results in terms of 
the $\langle \beta_T\rangle$ dependence on $\sqrt{\frac{dN}{dy}/S_{\perp}^{geom}}$ is
presented in Fig.10.  One should mention that for the BES energies \cite{STAR2} 
the BGBW fits were performed simultaneously on particles and antiparticles 
$p_T$ spectra, although they do not present the same trends in many 
respects. Therefore, in Fig.10, 
the $\langle \beta_T\rangle$ for antiparticles for some energies and centralities, where the azimuthal dependent BGBW fits were published \cite{Sun,Adam}, were represented by open symblols.
One could observe that, with increasing collision energy, the
values of $\langle \beta_T\rangle$ for antiparticles converge towards the values 
obtained from a simultaneous fit of particles and antiparticles $p_T$
spectra \cite{STAR1, STAR2}.  
However, the $\langle \beta_T\rangle$ values reported in the literature, scale rather nice as a function of $\sqrt{\frac{dN}{dy}/S_{\perp}^{geom}}$ and a $4^{th}$ order polynomial function
fits them well.
The fit quality can be followed in the bottom plot of Fig.10. 
Within the experimental error bars, all data follow the fit
result, except the points corresponding to the peripheral
collisions at the lowest BES energies. The fit parameters are included in the figure.
\begin{figure}[t!]
\includegraphics[width=0.95\linewidth]{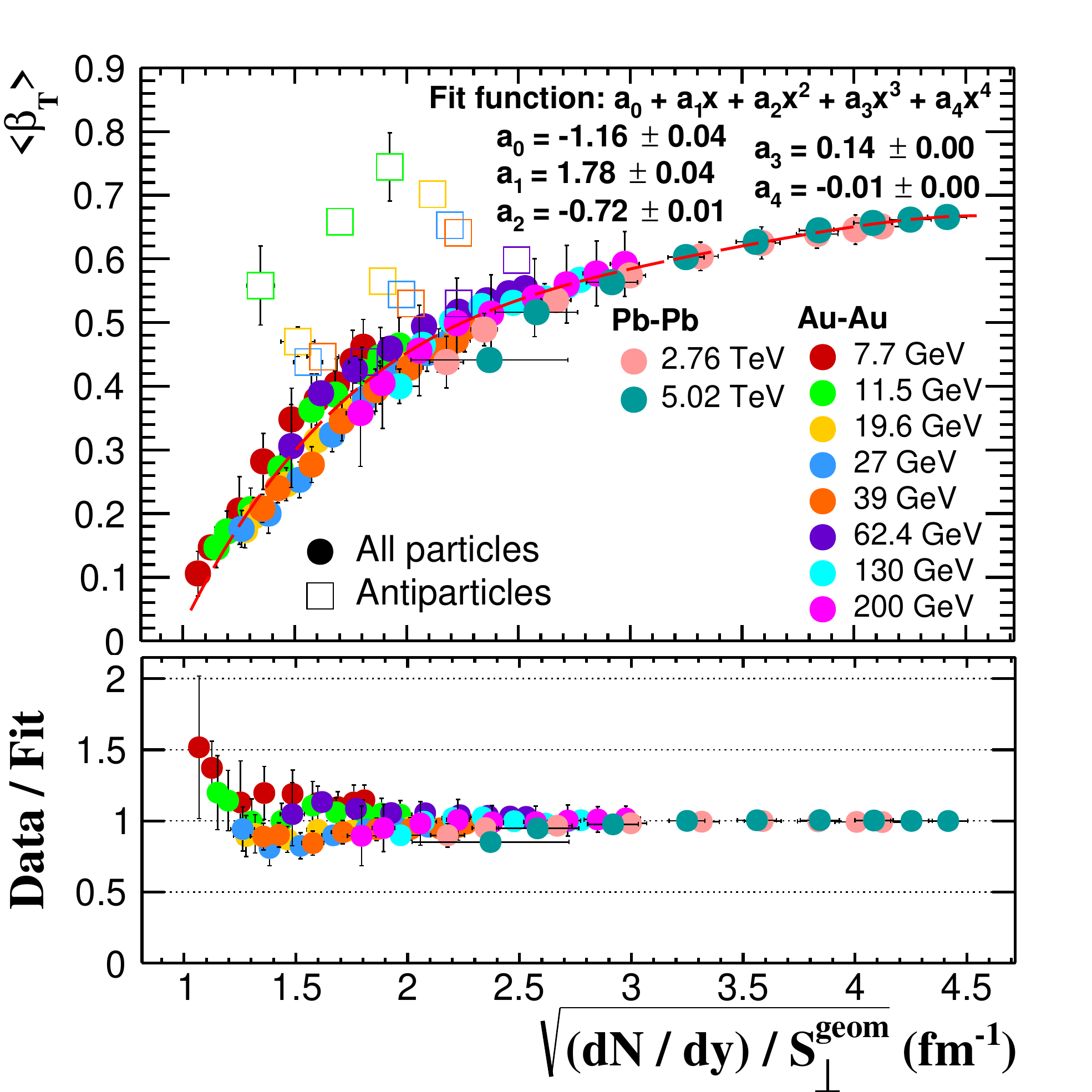}
\caption{Top: the BGBW fit parameter $\langle \beta_T\rangle$ as a function of$\sqrt{\frac{dN}{dy}/S_{\perp}^{geom}}$,
the dashed red line corresponds to the result of the fit using a 
$4^{th}$ order polynomial function. Bottom: the Data/Fit ratio.}
\label{fig-10}
\end{figure}
The same representation in which the data corresponding to $\sqrt{s_{NN}}$=7.7, 11.5 and 
62.4~GeV are excluded, can be followed in Fig.11. For the remaining energies, from 
$\sqrt{s_{NN}}$=19.6 GeV to 5.02~TeV a much better scaling is observed. 
\begin{figure}[]
\includegraphics[width=0.95\linewidth]{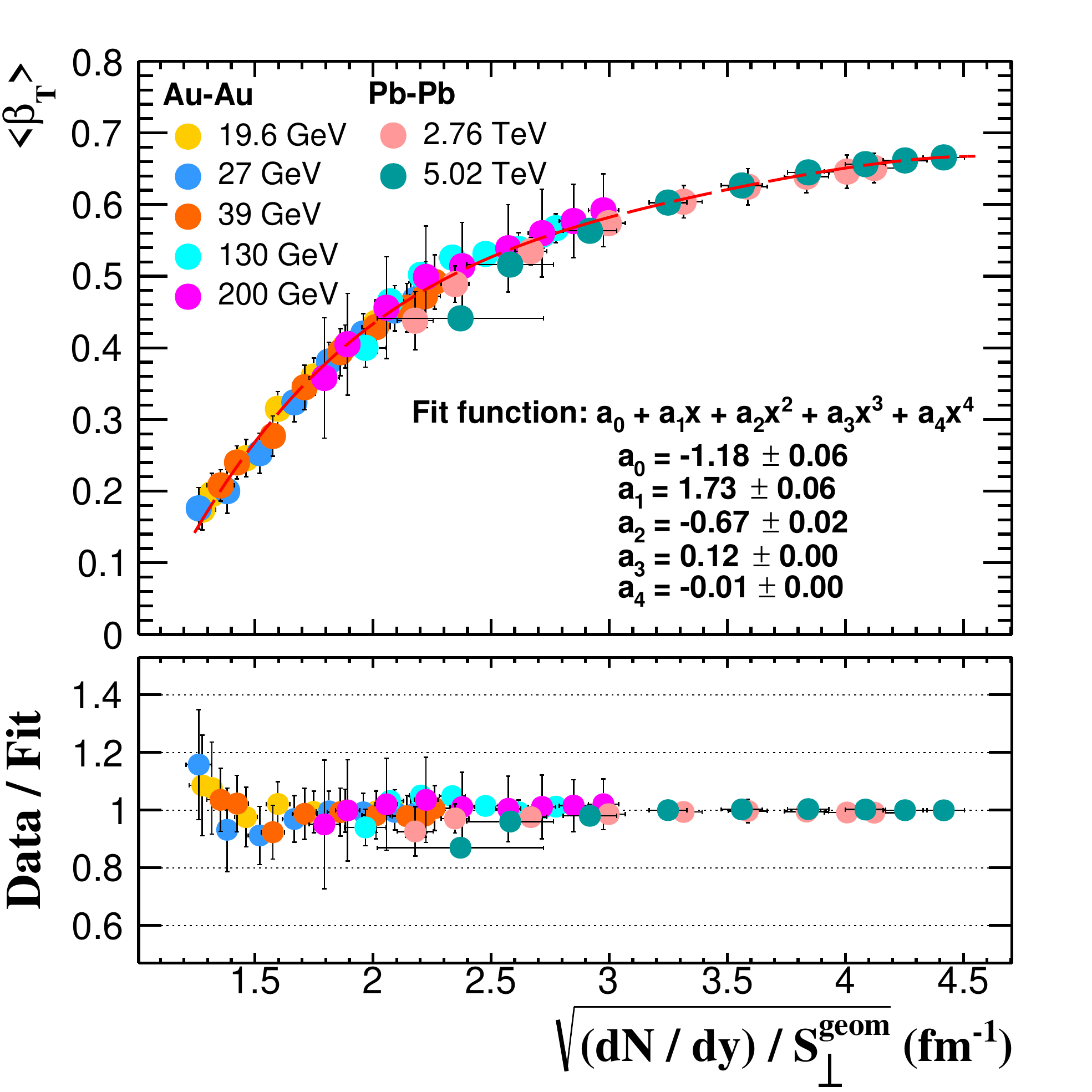}
\caption{The BGBW fit parameter $\langle \beta_T\rangle$ as a function of$\sqrt{\frac{dN}{dy}/S_{\perp}^{geom}}$ excluding $\sqrt{s_{NN}}$=7.7, 11.5 and 62.4 GeV 
collision energies.}
\label{fig-11}
\end{figure}
The dynamics in $\langle \beta_T\rangle$ as a function of 
$\sqrt{\frac{dN}{dy}/S_{\perp}^{geom}}$ for different collision energies can be easier followed in Fig.12 where the ratio between $\langle \beta_T\rangle$ at a given centrality relative to 
$\langle \beta_T\rangle$ in the most peripheral collisions, 70\%-80\% (58\%-85\% for 130 GeV), 
$\langle \beta_T\rangle$/$\langle \beta_T^{Peripheral} \rangle$, is plotted as a function of $\sqrt{\frac{dN}{dy}/S_{\perp}^{geom}}$ for all energies.

\begin{figure}[t!]
\includegraphics[width=1.\linewidth]{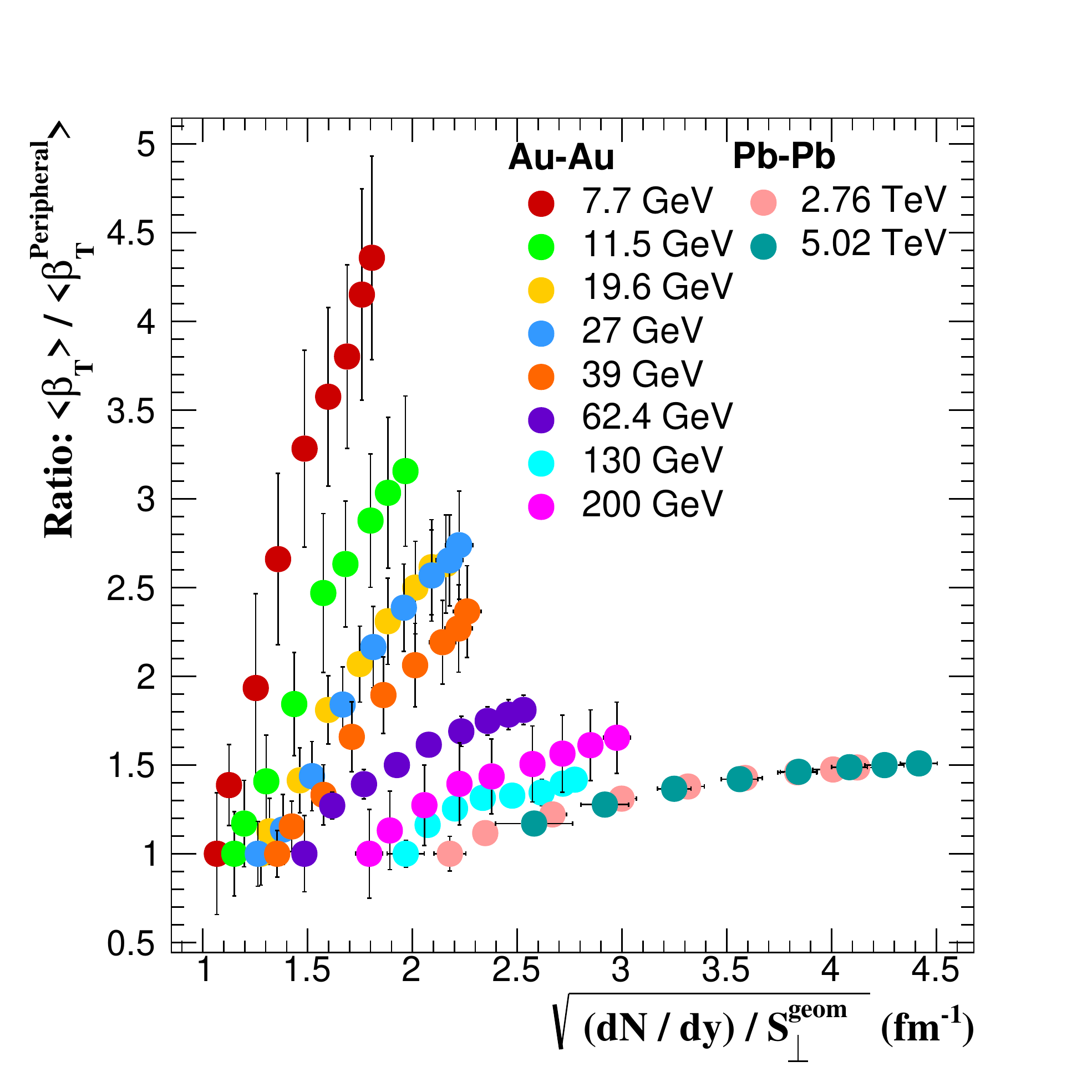}
\caption{$\langle \beta_T\rangle$/$\langle\beta_T^{Peripheral}\rangle$ as a function of$\sqrt{\frac{dN}{dy}/S_{\perp}^{geom}}$. $\langle \beta_T\rangle$ fit parameters were reported in Ref. \cite{STAR2, STAR1, ALICE1, ALICE2}.}
\label{fig-12}
\end{figure}
\noindent
In Fig.13 the $T_{kin}^{fo}$ and n parameters and their dependence on  
$\sqrt{\frac{dN}{dy}/S_{\perp}^{geom}}$ are presented.
A close to linear dependence with a negative slope is observed in Fig.13a, for $T_{kin}^{fo}$ at RHIC energies. Within the error bars, it is rather
difficult to conclude on some collision energy dependence of  
$T_{kin}^{fo}$ for a given value of the geometrical variable. 
\begin{figure}[t!]
\includegraphics[width=0.9\linewidth]{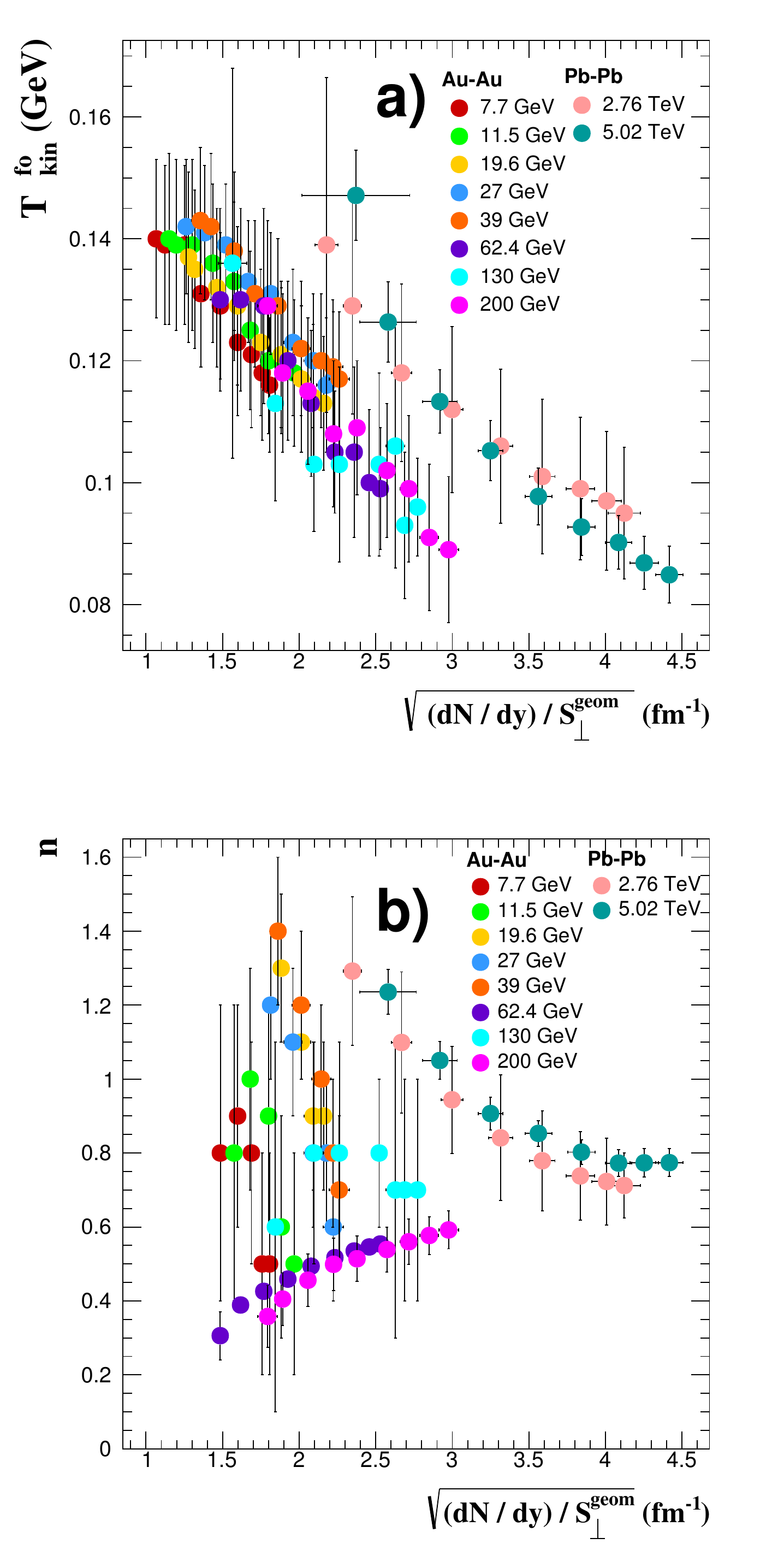}
\caption{a) $T_{kin}^{fo}$ and b) n parameters as a function of $\sqrt{\frac{dN}{dy}/S_{\perp}^{geom}}$. The fit parameters were taken from \cite{STAR2, STAR1, ALICE1, ALICE2, ALICE22}.}
\label{fig-13}
\end{figure}
On the other hand, a significant shift of about 20 MeV in $T_{kin}^{fo}$ 
fit parameter towards larger values is evidenced for a given 
$\sqrt{\frac{dN}{dy}/S_{\perp}^{geom}}$ at LHC energies relative to the RHIC energies. Similar shifts were mentioned in the previous chapters for 
$\langle p_T\rangle$ and the offsets of  
$\langle p_T\rangle$ as a function of mass.
Such a shift is also evidenced in the $T_{kin}^{fo}$ versus $\langle \beta_T\rangle$ representation in Fig.14 where the fit parameters reported in Ref. \cite{STAR2, STAR1, ALICE1, ALICE2, ALICE22} are used. 
\begin{figure}[t!]
\includegraphics[width=1.05\linewidth]{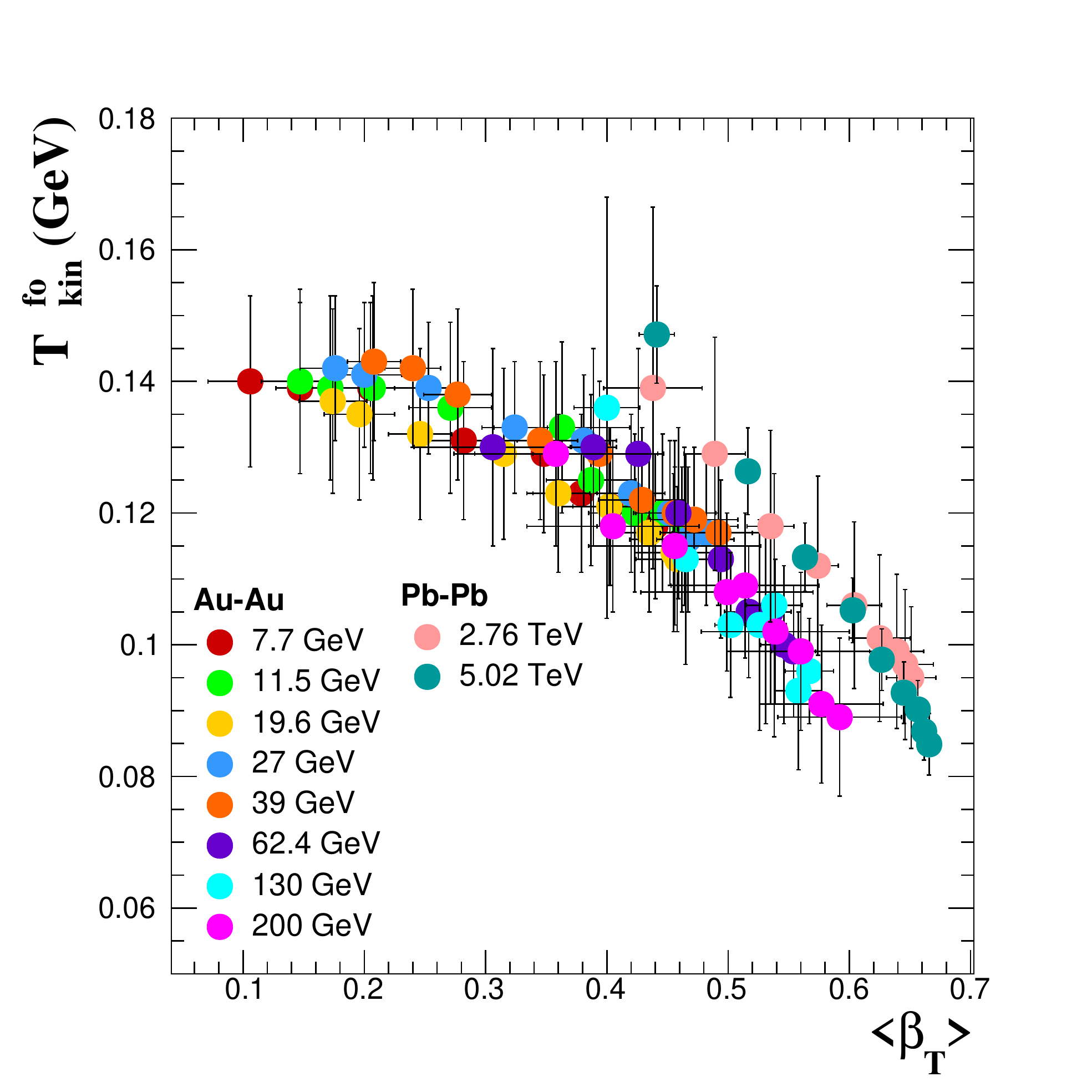}
\vspace{-.5cm}
\caption{The BGBW fit parameters $T_{kin}^{fo}$ versus $\langle \beta_T\rangle$ reported in \cite{STAR2, STAR1, ALICE1, ALICE2, ALICE22}.}
\label{fig-14}
\end{figure}
As far as the n dependence on $\sqrt{\frac{dN}{dy}/S_{\perp}^{geom}}$ is concerned, Fig.13b, the values for BES energies are rather scattered and those corresponding to 62.4
and 200 GeV show an opposite trend to what is observed at LHC. Usually, the flow 
profile changes from a shell type expansion, large n values, towards n=1 (Hubble type) with increasing centrality, even smaller than 1 for 
very central collisions. 
It is worth mentioning that for a consistent interpretation, the fits of the $p_T$ spectra  using the BGBW expression have to be
done at all energies on the same $p_T$ range for a given species. The range has to be chosen such to reduce as much as possible the influence of processes other than
collective expansion on the extracted fit parameters. 
Therefore, the lower limit of
the fit range for pions has to be chosen
such that the contribution coming from resonance decays is reduced, while the upper fit ranges
for all species 
have to be optimised in order to 
be influenced as little as possible by the suppression effects. Last but not least, the
influence of the corona contribution on the fit parameters has to be carefully 
considered.
\section{Comparison between $pp$ and $Pb-Pb$ systems at LHC energies}
 Similarities between pp and Pb-Pb in terms of the behaviour of different observables, like 
the ($\langle\beta_T\rangle$ - $T_{kin}^{fo}$) correlation as a function of charged particle multiplicity \cite{Cristi1} and near-side long range pseudorapidity correlations at
large charged particle multiplicities \cite{CMS1}, were evidenced
at LHC energies. 
The extent to which the similarity between pp and \mbox{Pb-Pb}
is also evidenced in the behaviour of the 
observables described in the previous chapters as a function of the saturation momentum, i.e.  
$\sqrt{\frac{dN}{dy}/S_{\perp}}$, is further investigated.
 For this comparison we used the results of the ALICE Collaboration for $p_T$ spectra 
of identified light flavour charged hadrons as a function of charged particle multiplicity at 
mid-rapidity as well as the results of their fits with the BGBW expression given by 
Eq.5 \cite{Cristi1}. The hadron density per unit of rapidity for the 
mid-central charged particle multiplicity was estimated by extrapolating 
the results reported by the ALICE Collaboration in Ref. \cite{ALICE10}.
The $\langle p_T\rangle$
values were estimated based on the $p_T$ spectra from 
\cite{Cristi1} extrapolated in the unmeasured regions using fits of the 
measured spectra with the expression from \cite{Byl}:
\begin{equation}
\frac{d\sigma}{p_Tdp_T}=A_eexp\left(-E_T^{kin}/T_e\right)+\frac{A}{\left(1+\frac{p_T^2}{T^2\cdot n}\right)^n}
\end{equation}
The interaction area for pp collisions, $S_{\perp}^{pp}$=$\pi$$R_{pp}^2$, is calculated using the estimates of the maximal radius for 
which the energy 
density of the Yang-Mill fields is larger than 
$\varepsilon=\alpha\Lambda_{QCD}^4$ ($\alpha\in{[1,10]}$) 
within the IP-Glasma initial state model \cite{Bazd1, Sch1}.
Within the present knowledge of QCD, $\alpha$ cannot be precisely 
estimated.
The $r_{max}$ values used in Ref. \cite{Bazd1} for $\alpha$=1
were fitted in Ref. \cite{McL2} with the following 
expressions: 
\begin{equation}
f_{pp}=
\left\{
\begin{array}{rl}
0.387+0.0335x+0.274x^2-0.0542x^3 & \mbox{if $x<3.4$}\\
1.538                            & \mbox{if $x\geq$ 3.4} 
\end{array}
\right.
\end{equation}
Using the same recipe we fitted the $r_{max}$ values from Ref. \cite{Bazd1} for $\alpha$=10 with the following expression:
\begin{equation}
f_{pp}=
\left\{
\begin{array}{rl}
-0.018+0.3976x+0.095x^2-0.028x^3 & \mbox{if $x<3.4$}\\
1.17                            & \mbox{if $x\geq$ 3.4} 
\end{array}
\right.
\end{equation}
where x=$(dN_g/dy)^{1/3}$. 
The hadron density per unit of rapidity, estimated based on the following approximation:
$\frac{dN}{dy}\simeq \frac{3}{2}\frac{dN}{dy}^{(\pi^+ + \pi^-)}+ 2\frac{dN}{dy}^{(p+\bar{p}, \Xi^-+\bar{\Xi^+}, K^0_S)}+
\frac{dN}{dy}^{(K^++K^-,\Lambda + \bar{\Lambda}, \Omega^- + \bar{\Omega}^+)}$
 and the corresponding overlapping areas
for $\alpha$=1 and $\alpha$=10 values are listed in Table VI.
\begin{table}
\begin{tabular}{| c | c | c | c |}
\hline
$\sqrt{s}$ (TeV)  & 
dN/dy & 
\multicolumn{2}{| c |}{$S_{\perp}$ ($fm^{2})$} \\ \cline{3-4} 
(pp) & & $\alpha$ = 1 & $\alpha$ = 10 \\ \hline
&82.1$\pm$2.8 & 7.43$\pm$0.48 &  4.30$\pm$0.36 \\ \cline{2-4} & 
70.2$\pm$2.2 & 7.43$\pm$0.41 &  4.30$\pm$0.31 \\ \cline{2-4} &
59.4$\pm$1.7 & 7.43$\pm$0.35 &  4.30$\pm$0.27 \\ \cline{2-4} 7&
48.8$\pm$1.3 & 7.43$\pm$0.30 &  4.30$\pm$0.23 \\ \cline{2-4} &
37.3$\pm$0.9 & 7.39$\pm$0.02 &  4.20$\pm$0.02 \\ \cline{2-4} &
26.8$\pm$0.6 & 6.89$\pm$0.05 &  3.80$\pm$0.03 \\ \cline{2-4} &
18.2$\pm$0.4 & 5.94$\pm$0.06 &  3.16$\pm$0.04 \\ \cline{2-4} &
10.8$\pm$0.2 & 4.58$\pm$0.06 &  2.29$\pm$0.04 \\
\hline
\end{tabular}
\caption{The hadron density per unit of rapidity and transverse overlapping areas for $\alpha$=1 and 
$\alpha$=10 for pp collisions at $\sqrt{s}$=7 TeV.}
\end{table}
The gluon density per unit of 
rapidity was approximated by 
$dN_{g}/dy\approx dN/dy$.   
\begin{figure}[b!]
\includegraphics[width=1.\linewidth]{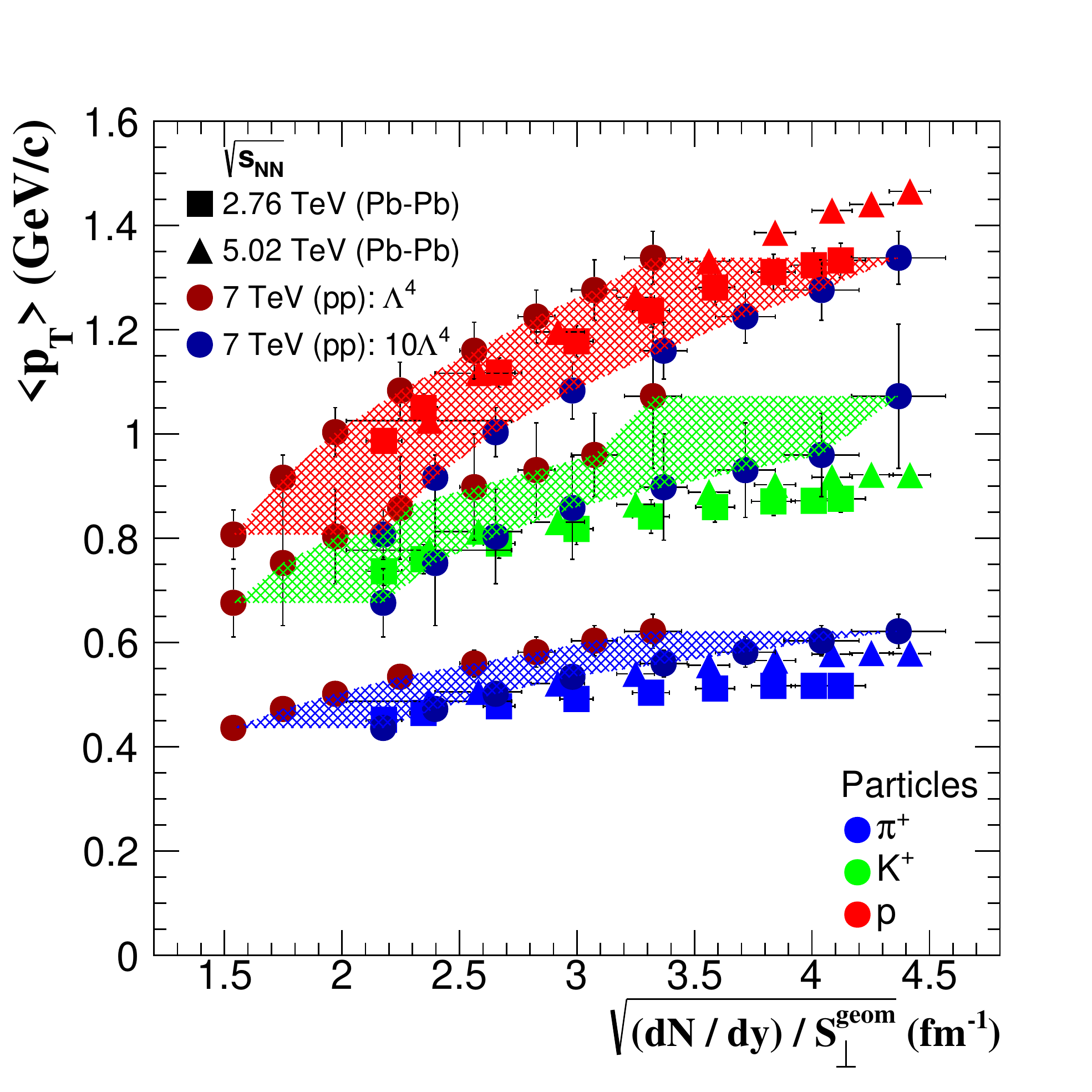}
\caption{$\langle p_T\rangle$
for identified charged hadrons for pp collisions at $\sqrt{s}$=7 TeV \cite{Cristi1}
(dark red symbols - 
$\alpha$=1, dark blue symbols - $\alpha$=10)
and Pb-Pb
collisions at $\sqrt{s_{NN}}$=2.76 and 5.02~TeV  \cite{ALICE1, ALICE2}
The blue, green and red shaded areas 
represent the uncertainty in the overlapping surface area estimates for 
pp collisions, see the text.
}
\label{fig-15}
\end{figure}
\begin{figure}[t!]
\includegraphics[width=1.\linewidth]{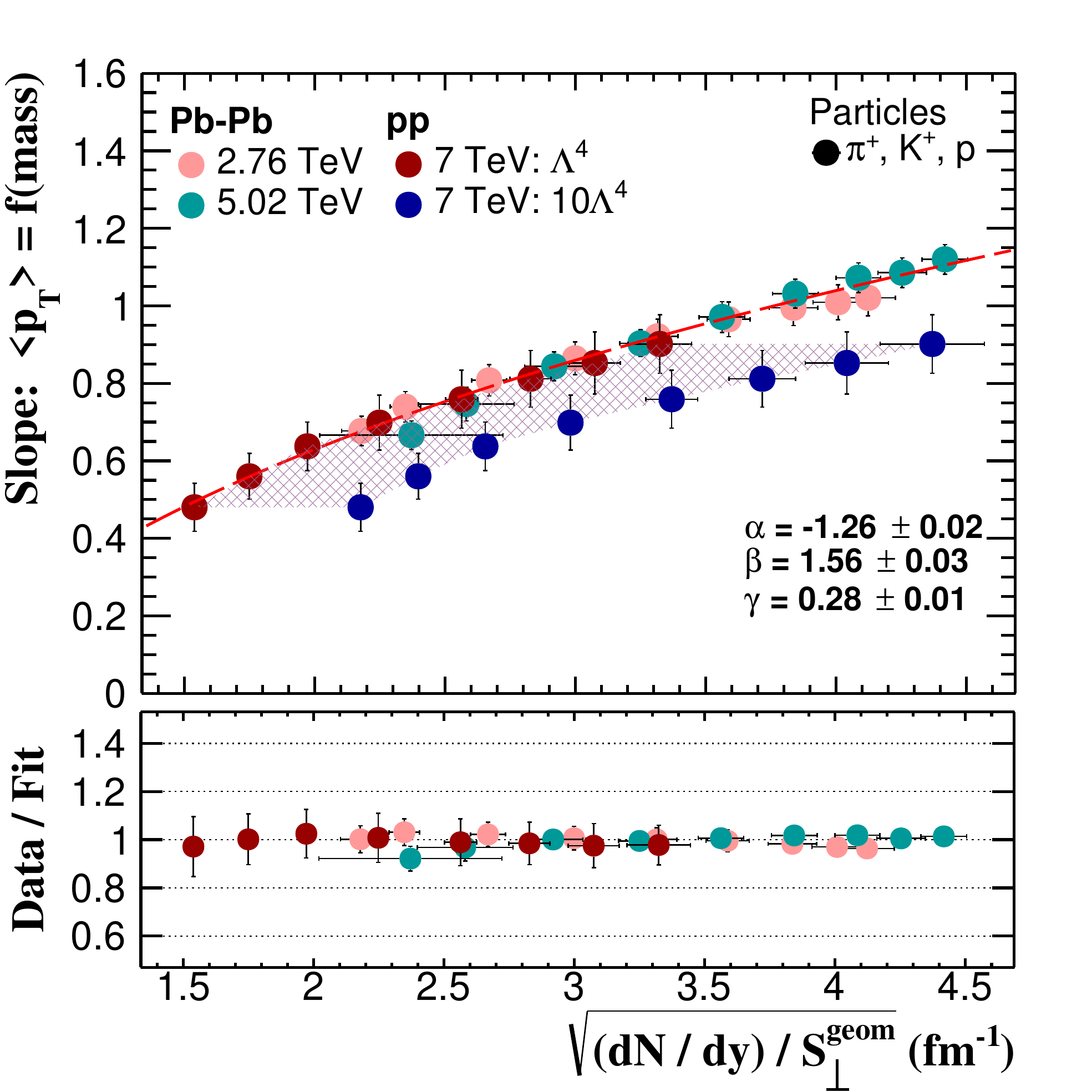}
\caption{The slopes of the $\langle p_T\rangle$ particle mass dependence as a function of$\sqrt{\frac{dN}{dy}/S_{\perp}^{geom}}$ for pp at $\sqrt{s}$=7 TeV (red symbols - 
$\alpha$=1, blue symbols - $\alpha$=10) and Pb-Pb at $\sqrt{s_{NN}}$=2.76 and 5.02 TeV. 
}
\label{fig-16}
\end{figure}
 The comparison between the $\langle p_T\rangle$ dependence on the 
square root of the hadron density per unit of rapidity and unit of interaction area for 
the pp 
at $\sqrt{s}$=7 TeV and Pb-Pb at $\sqrt{s_{NN}}$=2.76 and 5.02~TeV 
collisions, based on the results 
obtained by the ALICE Collaboration \cite{Cristi1, ALICE1, ALICE2, ALICE22}, is 
presented in Fig.15. As one could see, the general trend for all the three species 
is very similar in pp and Pb-Pb collisions. The differences could have 
several origins, i.e. the difference in the collision energies, a
systematic larger 
$\langle p_T\rangle$ for kaons in pp relative to Pb-Pb, uncertainty in 
estimating the value of $\alpha$, the large inhomogeneity of the initial state with a direct consequence on the $S_{\perp}$ estimate and last but 
not least the build up of collective expansion in the hadronic phase and
suppression effects taking place in the Pb-Pb case and not yet evidenced 
in pp collisions.   
\begin{figure}[t!]
\includegraphics[width=1.\linewidth]{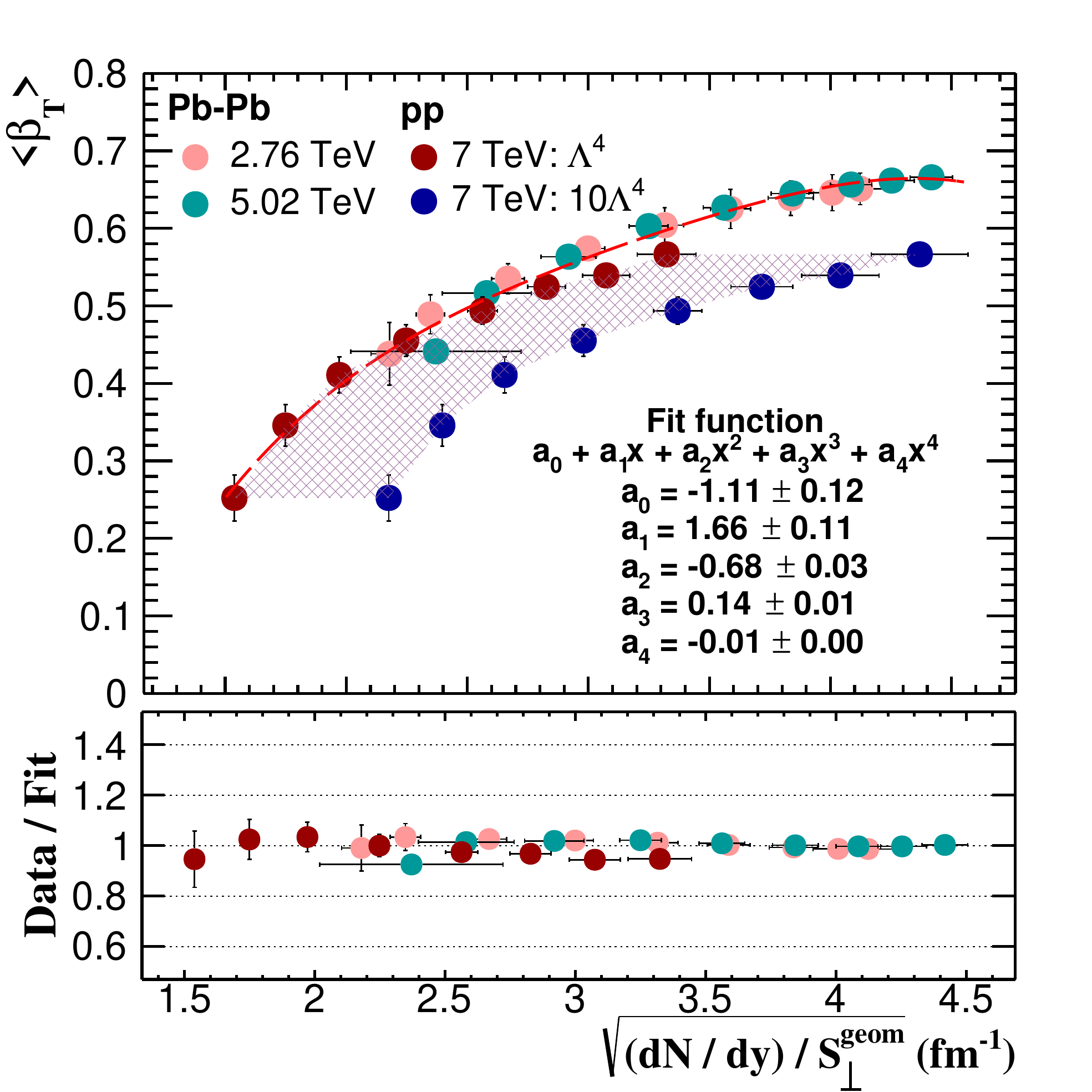}
\caption{The BGBW fit parameters $\langle \beta_T\rangle$ as a function of
$\sqrt{\frac{dN}{dy}/S_{\perp}^{geom}}$ for pp at $\sqrt{s}$=7 TeV and Pb-Pb at 
$\sqrt{s_{NN}}$=2.76 and 5.02 TeV.
The 
shaded area  
represents the uncertainty in the overlapping area estimates for 
pp collisions, see the text.}
\label{fig-17}
\end{figure}
The comparison 
between the two systems in terms of 
the slopes of the $\langle p_T\rangle$ particle mass dependence as a function of $\sqrt{\frac{dN}{dy}/S_{\perp}^{geom}}$ 
is presented in
Fig.16. A very good scaling is found using $\alpha$=1 for pp collisions.
The same value of $\alpha$ was used in Refs.\cite{McL2, McL3}. These results seem to support the assumption that the global properties of the hadron production are 
determined by the properties of flux tubes of 
$\sim$1/$\sqrt{\frac{dN}{dy}/S_{\perp}}$ size and are very little influenced 
by the size of the colliding system \cite{Bir,Shu,McL2}. 
A similar behaviour was evidenced at the baryonic level at much lower 
energies  where the main features of the dynamic evolution of 
the fireball are determined by the initial baryon density profile and 
temperature and not too much by its size \cite{Pet2}. 
As it is well known, the LPHD
approach neglects all collective effects. However, a comparison 
between pp and Pb-Pb collisions in terms of $\langle \beta_T\rangle$, one of the BGBW fit 
parameters interpreted as the average transverse flow velocity, could be rather interesting. 
$\langle \beta_T\rangle$ values for pp at $\sqrt{s}$=7 TeV \cite{Cristi1} and Pb-Pb at $\sqrt{s_{NN}}$=2.76 and 5.02 TeV \cite{ALICE1, ALICE2, ALICE22} reported by the ALICE Collaboration are 
represented as a function of $\sqrt{\frac{dN}{dy}/S_{\perp}^{geom}}$ in Fig.17. A $4^{rd}$ degree polynomial function fits rather well the data corresponding
to pp at $\sqrt{s}$=7 TeV ($\alpha$=1) and Pb-Pb at $\sqrt{s_{NN}}$=2.76 and 5.02 TeV. The fit quality 
is represented in the bottom plot of the figure. Qualitatively the trends are similar and there is even a
very good quantitative scaling for $\alpha$=1 used in the estimate of
$S_{\perp}$ for the pp case. The origin of the remaining differences was
discussed above. This similarity shows that the main features of the 
dynamical evolution of the systems produced in pp or Pb-Pb collisions at 
LHC energies are determined by the density of produced hadrons per unit of 
rapidity and overlapping area.

\section{Conclusions} 

   Based on the data for the highest three energies measured at RHIC
($\sqrt{s_{NN}}$=62.4, 130, 200 GeV), the most recent results from BES at RHIC
($\sqrt{s_{NN}}$=7.7-39 GeV)
and the highest
collision energies at LHC ($\sqrt{s_{NN}}$=2.76, 5.02 TeV), we performed a systematic study of the dependence of different observables on the 
geometrical variable calculated as the
square root of the 
hadron density per unit of rapidity and unit of overlapping area of two colliding ions.
The overlapping area has been estimated in the Glauber MC approach.
The experimental $\langle p_T\rangle$ values follow a rather good scaling as a function of 
this variable for each energy.
Linear fits of the experimental data show slopes which increase from pions to 
protons and decrease from BES to LHC energies. 
A saturation trend  for the most central collisions at LHC is observed. 
For $\sqrt{s_{NN}}$=200 GeV, 2.76 TeV and 5.02 TeV the $\langle p_T\rangle^{core}$ and 
$\sqrt{\frac{dN}{dy}^{core}/(S_{\perp}^{geom})^{core}}$
were estimated based on the core-corona approach. The corresponding 
$\langle p_T\rangle^{core}$
versus 
$\sqrt{\frac{dN}{dy}^{core}/(S_{\perp}^{geom})^{core}}$ show lower slopes and their decrease going from $\sqrt{s_{NN}}$=200 GeV
to 5.02 TeV is less evident for all three species. 
   This shows the importance of discriminating between the corona and core contributions in such 
a type of analysis, for a quantitative comparison.
   The decrease in the slopes from RHIC to LHC for all species and for the most central collisions at 
LHC energies seems to support the approach presented in Ref. \cite{Lev1}. 
A much better scaling as a function of 
$\sqrt{\frac{dN}{dy}/S_{\perp}^{geom}}$ is observed 
for the slope from the linear fit of the $\langle p_T\rangle$ dependence on the particle mass
and  the BGBW fit parameter, $\langle \beta_T\rangle$.
The offset of the $\langle p_T\rangle$ particle mass dependence and the $T_{kin}^{fo}$ parameter show a clear jump towards larger values between RHIC and LHC energies.
   As it was already mentioned, other phenomena, like suppression and its azimuthal dependence as well as
the hydrodynamic expansion in the deconfined and after hadronization stages,  
also have to be considered.  
  The very similar dependence of the $\langle p_T\rangle$, $\langle p_T\rangle$ 
  particle mass dependence and  the BGBW fit parameter, $\langle \beta_T\rangle$, on 
$\sqrt{\frac{dN}{dy}/S_{\perp}}$ in pp and Pb-Pb collisions at LHC energies support the assumption that the global properties evidenced at LHC energies are 
determined by the properties of flux tubes of 
$\sim$1/$\sqrt{\frac{dN}{dy}/S_{\perp}}$ size, the system size playing a 
minor role.
\begin{acknowledgments}
This work was carried out under the contracts sponsored by the Ministry of Research and Innovation: 
RONIPALICE-04/16.03.2016 (via IFA Coordinating Agency) and
PN-18 09 01 03.
\end{acknowledgments}

\bibliographystyle{unsrt}

\end{document}